\documentstyle[11pt,epsfig,graphicx]{article}
 
\newcommand \beq{\begin{eqnarray}}
\newcommand \eeq{\end{eqnarray}}
 
\begin{document}
 
\title{{\sf Neutron stars in a Skyrme model with hyperons} \\
\author{L. Mornas\footnote{lysiane@pinon.ccu.uniovi.es}  \\
{\small {\it Departamento de F{\'\i}sica, Universidad de Oviedo,
Avda Calvo Sotelo 18, 33007 Oviedo, Spain}}}          
}
\maketitle

\begin{abstract}
Available Skyrme parametrizations with hyperons are examined
from the point of view of their suitability for applications
to neutron stars. It is shown that the hyperons can attenuate or 
even remove the problem of ferromagnetic instability common to 
(nearly) all Skyrme parametrizations of the nucleon-nucleon 
interaction. At high density the results are very sensitive to 
the choice of the $\Lambda$-$\Lambda$ interaction. The selected
parameter sets are then used to obtain the resulting properties
of both cold neutron stars and hot protoneutron stars. The 
general features known from other models are recovered.
\end{abstract}    

\vskip -0.5cm 
\null\hskip 0.33cm {\small PACS numbers:  97.60.Jd, 21.65.+f, 21.30.-x}
\vskip -2pt \hskip 0.33cm  {\small keywords: neutron stars, Skyrme 
forces, hyperons}

\section{Introduction}

Skyrme parametrizations provide a simple tool for calculating the
equation of state. Skyrme or Skyrme-inspired  models are often 
used in applications to neutron stars with adapted versions which 
pay special attention to the asymmetry properties. The Skyrme Lyon 
series \cite{Sly} is the most modern example of this approach and 
was designed to reproduce saturation properties, the main properties 
of nuclei as well as the results of simulations of pure neutron 
matter.

In neutron stars in chemical equilibrium, hyperons appear at a
threshold density about twice the nuclear saturation density.
A sizable hyperon fraction can be expected to be present in the core 
of the most massive neutron stars; the central density of the Skyrme
Lyon series for example is of the order of seven times saturation 
density in the star with maximum mass and three to four times 
saturation density in a 1.4 $M_\odot$ star.

Original Skyrme parametrizations did not take hyperons into account.
Balberg and Gal \cite{BG97} have remedied the fact with their 
parametrization of the energy density. It could be argued however 
that the description of the nucleon sector provided by this 
parametrization is rather poor. 

Rather than designing a new force in the hyperonic sector with the 
purpose of describing neutron stars, an other possibility would be 
to restrict oneself to the numerous available data sets which were 
fitted to nuclear properties. This is the approach taken by Rikovska 
Stone {\it et al.} \cite{RSMKSS03}, who recently tested 87 parameter 
sets for the NN force in neutron star calculations. 

In the hyperonic sector there exist a few parametrizations of 
the Lambda-Nucleon and Lambda-Lambda interaction fitted to
the properties of hypernuclei, including several well tested
sets by Lanskoy {\it el al.} \cite{LY97,L98} and an earlier set by 
Fernandez {\it et al.} \cite{FLAP89}. The present work can be considered
as an extension of the study of Rikovska Stone {\it et al.} including
the $\Lambda$ hyperon.

\medskip

A general feature of Skyrme models in the np sector is that they show
a ferromagnetic transition at rather low density \cite{VNB84,MNvG02}. 
If present, such a transition, apart from modifying the equation of state 
and chemical equilibrium, could give rise to induced magnetic fields in
rotating neutron stars. It also plays a crucial role in calculations of 
the neutrino mean free path in supernova and protoneutron stars.
Calculations of this parameter in non relativistic (Skyrme or Gogny)
models of nuclear matter \cite{NHV99,RPLP99,M01} thus obtained 
that the cross section diverges and the mean free path drops to zero 
at the transition.

It remains an open issue whether this transition is genuine or an 
artefact of the Skyrme model. While a free Fermi gas eventually
becomes ferromagnetic, the nuclear correlations are known 
to play a crucial role. Most non relativistic calculations which
take correlations into account, {\it e.g.} by solving the 
Brueckner-Hartree-Fock equations with modern bare NN potentials
\cite{VBPR02,ZLS02}, concluded that spin ordered matter was not favored 
energetically. Relativistic mean field models predict no ferromagnetic 
transition; on the other hand relativistic calculations in the Hartree-Fock 
approximation by Bernardos {\it et al.} \cite{BMNQ95} find a transition, 
albeit as a rather large density ($4\ n_{\rm sat}$). A recent work by
Maruyama and Tatsumi \cite{MT01}, although not putting forward 
a quantitative prediction, reaches a similar conclusion as to the 
importance of the Fock contribution.

In this work we will take the viewpoint that the ferromagnetic transition, 
especially when appearing at such low densities as 2--3 times saturation, 
is probably an artefact of the Skyrme model, and will select among the 
available Skyrme parametrizations those which as far as possible avoid or
delay the transition to higher densities. 

\medskip

This work includes the effects of temperature and neutrino trapping
and can also be applied in hot protoneutron stars. 

A further motivation of the present study was to obtain a model which 
would allow to calculate the neutrino-baryon scattering rate in the hot 
protoneutron star formed shortly after the supernova collapse. It was
therefore necessary to be able to calculate the Landau parameters in 
the spin $S=1$ channel. Besides determining the position of an eventual 
ferromagnetic transition, the Landau parameters in the spin $S=1$ channel 
play a central role when calculating the axial response function in the 
random phase approximation, which is the dominant contribution to the 
neutrino-baryon scattering rate. This application is presented in a 
separate paper \cite{M04b}.

\medskip

After a presentation of the available Skyrme parametrizations in section 
\S \ref{S:IntroSky}, the following sections proceed to select a few among 
them according to their suitability for neutron star applications. Section 
\S \ref{S:threshold} first examines the threshold density for hyperon 
formation under the conditions of $\beta$ equilibrium. \S \ref{S:ferro} 
presents a discussion of the issue of ferromagnetism. The selected parameter 
sets are used in section \S \ref{S:eos-MvsR} to obtain the properties of the 
corresponding neutron stars. The effect of non vanishing temperature and 
of a trapped neutrino fraction are studied in \S \ref{S:finiteT}. The main 
results are collected and discussed in the conclusion.

The formalism is kept to a minimum in the main text, while all
relevant formulaes are gathered in the Appendix.

\section{Skyrme parametrizations of the hyperonic sector}
\label{S:IntroSky} 

The usual Skyrme model of nuclear matter introduces the nucleon-nucleon 
potential
\beq
V_{NN}(r_1 -r_2) &=&  t_0\, (1+x_0 P_\sigma) \delta(r_1-r_2)
+ {1 \over 2} t_1\, (1+x_1 P_\sigma) \left[ k'{}^2\  \delta(r_1-r_2)
+ \delta(r_1-r_2)\ k^2 \right] \nonumber \\
&& + t_2\, (1+x_2 P_\sigma)  k'\, \delta(r_1-r_2)\, k + {1 \over 6} t_3\,
(1+x_3 P_\sigma)\, \rho_N^\alpha \left({r_1+r_2 \over 2}\right)
\delta(r_1- r_2) 
\eeq   
It can readily be generalized to include nucleon-Lambda and
Lambda-Lambda interaction potentials (see {\it e.g.} \cite{LY97,L98})
\beq
V_{N\Lambda} (r_N-r_\Lambda) &=&
u_0\, (1+y_0 P_\sigma) \delta(r_N-r_\Lambda) + {1 \over 2} u_1\,
\left[ k'{}^2\  \delta(r_N-r_\Lambda) +  \delta(r_N-r_\Lambda)\ k^2 \right]
\nonumber \\
&& + u_2\, k'\, \delta(r_N-r_\Lambda)\, k + {3 \over 8} u_3\, (1+y_3 P_\sigma)
\rho_N^\beta \left({r_N+r_\Lambda \over 2}\right)  \delta(r_N-r_\Lambda) \\
V_{\Lambda\Lambda} (r_1-r_2) &=& \lambda_0 \delta(r_1-r_2) +{1 \over 2}
\lambda_1 \left[ k'{}^2\ \delta(r_1-r_2) + \delta(r_1-r_2)\ k^2 \right]
\nonumber \\
&& + \lambda_2\, k'\, \delta(r_1- r_2)\, k + \lambda_3\, \rho_\Lambda
\rho_N^\gamma
\eeq   
The potentials normally also include spin orbit contributions. They
are not explicited here since they will not be used in the remainder of 
this paper.

The energy density is then obtained in the Hartree-Fock approximation
from 
\beq
{\cal E} &=& <\psi\, |\, H \,|\, \psi > \quad {\rm with} \ H=
\sum_{A=N,\Lambda} T_A  + {1 \over 2} \sum_{A,B=N,\Lambda} V_{AB} \nonumber \\ 
         &=& {\cal E}_{NN} + {\cal E}_{N\Lambda} + {\cal E}_{\Lambda\Lambda}
\eeq
In homogeneous matter the wave functions are formed from antisymmetrized
plane wave states. The explicit expression of the energy density is given 
in the Appendix.

In our search for some suitable sets of Skyrme parametrizations we have
tested 43 NN Skyrme forces in combination with 13 parametrizations 
of the N$\Lambda$ and 4 options for the $\Lambda\Lambda$ forces. 
The NN forces were chosen among: SIII, SkM*, SI', SIII', SV, RATP, SGI, 
SGII, SLy230a, SLy2, SLy4, SLy5, SLy6, SLy7, SLy9, SLy10, SkO, SkO', SkS3, 
SkI1, SkI2, SkI3, SkI4, SkI5, Rs, Gs, SkT4, SkT5, T6, SkP, Ska, MSka, SK272,
SK255, Skz-0, Skz-1, Skz-2, Skz-3, Skz-4, SkSc4, SkSc15, MSk7, SKX.
After demanding that these sets fulfill various constraints that will
be discussed in detail in the next two sections, only four NN forces
were sorted out: the SLy10 force from the Skyrme-Lyon series \cite{Sly},
the modern SkI3 and SkI5 forces by Reinhard and Flocard \cite{RF95}
and the older SV force by Beiner {\it et al.} \cite{BFvGQ75}.

For the nucleon-$\Lambda$ interaction we have tried several parameter 
sets given by Lanskoy and Yamamoto \cite{LY97,YBZ88} and by Fernandez 
{\it et al.} \cite{FLAP89}: The $N\Lambda$ sets numbered from I to V
in \cite{LY97} are named LYI-LYV here, the sets numbered from 1 to 6
in \cite{YBZ88} are named YBZ1-YBZ6, other choices are the SkSH1 and
SkSH2 sets of \cite{FLAP89} or to switch off the $N\Lambda$ interactions. 
Finally, Lanskoy gives in \cite{L98} three sets SLL1, SLL2, SLL3 for the
$\Lambda\Lambda$ interaction, or we can switch it off.

The parametrization of Lanskoy and Yamamoto \cite{LY97} was extracted 
from G-matrix calculations \cite{Yamamoto-G} performed with the J\"ulich 
and Nijmegen potentials and were tested on hypernuclei. Other $N\Lambda$ 
and $\Lambda\Lambda$ interactions were fitted directly to hypernuclei data. 
The values given by Lanskoy and Yamamoto \cite{LY97} assume that the nucleon 
sector is parametrized by the SkM* or SIII interactions while the parameter 
set of the Salamanca group \cite{FLAP89} was used together with the SkS3 
interaction. Even though this is not fully consistent, we also used other 
more modern parametrizations in the nucleon sector like the Skyrme Lyon 
sets \cite{Sly} in order to investigate the role of the ferromagnetic 
transition (see section \S \ref{S:ferro})

Selected values of the parameter sets are listed in tables I, II 
and III. For other parametrizations we refer the reader to the exhaustive
table published by Rikovska Stone  {\it et al.} in the case of the NN forces 
and the papers of Lanskoy {\it et al.} and Fernandez {\it et al.} for the 
N$\Lambda$ and $\Lambda\Lambda$ forces.

\medskip

As this model only includes the $\Lambda$ hyperon, it appears to be less 
complete than the model of Balberg and Gal \cite{BG97} which includes all 
the hyperons. As mentioned in the introduction, we fixed our choice on the 
parametrization by Lanskoy {\it et al.} because these authors provide the 
two-particle potential rather than only the unpolarized energy density.  
This leaves open the possibility of going beyond the Hartree-Fock
approximation and to investigate the response functions of the matter 
at the RPA level.

It would be straightforward to extend the model to take into account
other hyperons such as the $\Sigma^-$. Data about the $\Sigma^-$ in
nuclear matter however is rather scarce and the author is not aware
of an equivalent Skyrme parametrization including the $\Sigma$ at
the same level of precision. One possibility would be to adjust
Skyrme parameters to reproduce Brueckner calculations {\it e.g.}
using density matrix expansion techniques. On the other hand
the neglect of the $\Sigma$ hyperon is sometimes justified from
the very lack of observation of $\Sigma$ hypernuclei, when this fact
is interpreted as indicating that the $\Sigma$-nucleon force is 
in fact repulsive. In that case  $\beta$-equilibrium equations
would predict that the threshold for $\Sigma$ hyperon formation
in neutron stars is shifted to very large densities (see {\it e.g.}
\cite{BG97,MFGJ95,SBG00,Dabrowski}).

We will use the model of Balberg and Gal \cite{BG97} and another
by Banik and Bandhyopadhyay \cite{BB00} in order to estimate the error 
committed by neglecting other hyperons and in particular the $\Sigma^-$. 

\begin{center}
\begin{tabular}{|l|c|c|c|c|c|c|c|c|c|}
\hline
Model & $\alpha$ & $t_0$ & $t_1$ & $t_2$ & $t_3$ & $x_0$ & $x_1$ & $x_2$ 
& $x_3$ \\
\hline
SLy10 & 1/6 & 2506.77 & 430.98 & -304.95 & 13826.41 & 1.0398 & -0.6745
      & -1.0  & 1.6833 \\
SkI3  & 1/4 & -1762.88 & 561.608 & -227.09 & 8106.2 & 0.3083 & -1.1722
      & -1.0907 & 1.2926 \\ 
SkI5  & 1/4 & -1772.91 & 550.84 & -126.685 & 8206.25 & -0.1171 & -1.3088
      & -1.0487 & 0.3410 \\     
SV    & 1  & -1248.3 & 970.6 & 107.2 & 0. & -0.17 & 0. & 0. & 1. \\   
\hline
\end{tabular}
\begin{flushleft}
{\bf Table I}: Skyrme parameters for the NN interaction \\
($t_0$ is given in MeV.fm${}^3$, $t_1$ and $t_2$ in  MeV.fm${}^5$, $t_3$ in
 MeV.fm${}^{3+3\alpha}$, the other parameters are adimensional)
\end{flushleft}
\end{center}
\vskip 0.2cm

\begin{center}
\begin{tabular}{|l|c|c|c|c|c|c|c|c|}
\hline
Model & $\beta$ & $u_0$ & $u_1$ & $u_2$ & $u_3$ & $y_0$ & $y_3$ & $V_\Lambda$\\
\hline
LY-I & 1/3 & -476. & 42. & 23. & 1514.1 & -0.0452 & -0.280 & -27.62 \\
LY-IV & 1/3 & -542.5 & 56.0 & 8.0 & 1387.9 & -0.1534 & 0.1074 & -28.17\\
YBZ-1 & 1 & -349.0 & 67.61 & 37.39 & 2000. & -0.108 & 0. & -26.52\\
YBZ-5 & 1 & -315.3 & 23.14 & -23.14 & 2000 & -0.109 & 0. &  -28.50\\
YBZ-6  & 1 & -372.2 & 100.4 & 79.60 & 2000. & -0.107 & 0. & -24.98\\
SkSH1 & -- & -176.5 & -35.8 & 44.1 & 0 & 0 & -- & -27.68 \\
\hline
\end{tabular}
\begin{flushleft}
{\bf Table II}: Skyrme parameters for the N$\Lambda$ interaction 
\cite{LY97,YBZ88} \\
($u_0$ is given in MeV.fm${}^3$, $u_1$ and $u_2$ in  MeV.fm${}^5$, $u_3$ in
 MeV.fm${}^{3+3\beta}$, the other parameters are adimensional. $V_\Lambda$
is the potential felt by a $\Lambda$ hyperon in nuclear matter at saturation
and is given in MeV.)
\end{flushleft}
\end{center}
\vskip 0.2cm

\begin{center}
\begin{tabular}{|l|c|c|}
\hline
 Model & $\lambda_0$ & $\lambda_1$ \\
\hline
SLL1 & -312.6 & 57.5 \\
SLL2 & -437.7 & 240.7 \\
SLL3 & -831.8 & 922.9 \\
\hline
\end{tabular} \\
\vskip 0.2cm
\begin{minipage}{10cm}
\begin{flushleft}
{\bf Table III}: Skyrme parameters for the $\Lambda\Lambda$ interaction 
\cite{L98} \\
($\lambda_0$ is given in MeV.fm${}^3$ and $\lambda_1$ in  MeV.fm${}^5$)
\end{flushleft}
\end{minipage}
\end{center}

\section{Threshold for hyperon formation}
\label{S:threshold}

In neutron star matter we impose that the conditions for $\beta$-equilibrium 
are fulfilled. The neutron, proton and Lambda hyperons are subject
to the processes
\beq
p + e^- \rightarrow n + \nu_e \quad ; \quad  p + e^- \rightarrow \Lambda 
+ \nu_e \quad\nonumber
\eeq
as well as their inverse
\beq
n \rightarrow p + e^- + \overline\nu_e \quad ; \quad
\Lambda \rightarrow p + e^- + \overline\nu_e
\eeq
We write the equality of the chemical potentials
\beq 
\hat\mu = (\mu_n + m_n) -(\mu_p + m_p) = \mu_e -\mu_\nu \quad ;
\mu_n + m_n = \mu_\Lambda + m_\Lambda
\label{eq:chemeq}
\eeq
We must also impose electric charge conservation
\beq
n_e = n_p
\label{eq:charge}
\eeq
At a given baryonic density the electron, proton and hyperon fractions 
are determined by the solution of Eqs. (\ref{eq:chemeq},\ref{eq:charge}).
The chemical potentials are determined by deriving the Skyrme energy density 
functional with respect to the density of corresponding particle. Their
explicit expression is given in the Appendix. The electrons are relativistic 
and their chemical potential is given by $\mu_e=\sqrt{k_{Fe}^2 + m_e^2}$.
In protoneutron star matter with trapped neutrinos we have $\mu_\nu = 
(6 \pi^2 n_\nu)^{(2/3)}$, while in colder neutrino-free neutron star matter
$\mu_\nu =0$. 

The equations should actually take into account the muons in the equation
for charge conservation together with the condition $\mu_e=\mu_\mu$. The
muons appear namely around saturation density. The muons were neglected in
this simplified model. Their effect is in fact not very important, especially
in our case where the charged $\Sigma^-$ is also absent from the model.

The threshold for $\Lambda$ hyperon formation is determined by the 
condition
\beq
\mu_\Lambda(thr) &=& \mu_\Lambda(k_{F\Lambda}=0) = \mu_n + m_n 
- m_\Lambda \nonumber \\
&=& u_0 \left(1 + {y_0\over 2} \right) \rho_N + {3 \over 8} u_3 
\left(1 + {y_3\over 2} \right) \rho_N^{\beta +1} + {1 \over 8} 
\left[ u_1 (2 + y_1) + u_2 (2 + y_2) \right] k \rho_N^{5/3}
\eeq
with $k =(3/5)\, (3 \pi^2)^{(2/3)}$. It depends on the strength 
of the $\Lambda$N force through the parameters $u_0, u_1, u_2, u_3..$ 
and indirectly on the $NN$ force through the chemical potential of 
the neutron (taken in $npe$ matter in $\beta$ equilibrium) and is 
independent of the $\Lambda$-$\Lambda$ force.
Depending on the choice of the parameters it is found that the 
threshold generally occurs between $1.7\ n_{\rm sat}$ and 
$4\ n_{\rm sat}$ in agreement with other Brueckner-Hartree-Fock 
or relativistic mean field calculations. When numerical values are 
inserted one can see that the value of $\mu_\Lambda(thr)$ is the 
result of a delicate cancellation between the $u_0$ and $u_3$ 
terms, the contribution of the $u_1,u_2$ term being of the order 
of the sum of the $u_0$ and $u_3$ terms. For example at $\rho_N
= 3\ n_{\rm sat}$ and $n_{\rm sat}=0.16\ {\rm fm}^{-3}$ we have 
for the LY-I parametrization of the N$\Lambda$ force
\beq
&& u_0 \left(1 + {y_0\over 2} \right) \rho_N =-223.32\ {\rm MeV} 
   \quad ,\quad
   {3 \over 8} u_3 \left(1 + {y_3\over 2} \right) \rho_N^{\beta +1} 
   =183.51\ {\rm MeV} \nonumber \\ 
&& {1 \over 8} \left[ u_1 (2 + y_1) + u_2 (2 + y_2) \right] k \rho_N^{5/3} 
   =27.46\ {\rm MeV} \nonumber
\eeq

Let us also note that the value of $\mu_\Lambda(k_{F\Lambda}=0)$
as saturation density with equal number of neutron and protons
is nothing but the single particle potential felt by a $\Lambda$ 
impurity in nuclear matter, {\it i.e.} it should equal to the
binding potential $V_\Lambda \simeq -28$ MeV obtained from data 
on hypernuclei. The actual values of $V_\Lambda=\mu_\Lambda(k_{F\Lambda}=0,
nB=0.16\ {\rm fm}^{-3})$ are reported in the last column of table II. 

We may conclude this section by stating that, once the parametrization
of the N$\Lambda$ force has been chosen so that the potential felt by a 
$\Lambda$ in nuclear matter reproduces the experimental value, the
condition that the threshold density for $\Lambda$ hyperons in
nuclear matter in $\beta$-equilibrium should lay around 2 -- 3 times 
saturation density does not severely constrain the admissible Skyrme 
NN and N$\Lambda$ forces. We observe that stiffer equations of state 
(eos) and the eos allowing for a larger proton fraction have lower hyperon 
thresholds (see table V for a sample of the results).

\section{Transition to ferromagnetic state}
\label{S:ferro}

\subsection{Criterion for a ferromagnetic instability}

Previous studies on the neutrino mean free path in neutron matter
\cite{NHV99} or $npe^-$ matter in $\beta$ equilibrium \cite{M01}
found that a pole appears in the calculation of the axial structure 
function above a certain critical density. This feature is typical
of Skyrme models and is related to a transition to a ferromagnetic 
state. In this section we will study how this critical density is 
affected by the presence of hyperons. 

Let us define the magnetic susceptibilities
$\chi_{ij}$ where $i,j \in \{ n,p,\Lambda \}$:
\beq
{1 \over \chi_{ij} }= {\partial^2 {\cal E} \over \partial {\cal M}_i 
\partial {\cal M}_j}\ , \qquad  {\cal M}_i=\mu_i (\rho_{i\uparrow}
-\rho_{i\downarrow}) 
\eeq
where ${\cal E}={\cal E}(\rho_{n\uparrow},\rho_{n\downarrow},
\rho_{p\uparrow},\rho_{p\downarrow},\rho_{\Lambda\uparrow},
\rho_{\Lambda\downarrow})$ is the polarized energy density functional,
${\cal M}_i$ are the magnetizations and $\mu_i$ are the magnetic moments.
The inverse susceptibilities are therefore proportional to the second 
derivatives of the energy density functional with respect to the 
polarizations:
\beq
{\mu_i \mu_j \over \chi_{ij} } &=& {2 \rho \over \rho_i \rho_j} \Delta_{ij} 
\ , \qquad
\Delta_{ij} = {1 \over 2} {\partial^2 ({\cal E}/\rho) \over \partial
s_i \partial s_j} \quad {\rm with}\ 
s_i = {\rho_{i\uparrow} -  \rho_{i\downarrow} \over
       \rho_{i\uparrow} +  \rho_{i\downarrow}}
\eeq
On the other hand it can be shown that the $\Delta_{ij}$ are related 
to the Landau parameters $g_0^{ij}$ defined in the Appendix through
\beq
\Delta_{ij} &=& {2 \rho \over \rho_i \rho_j} {1 \over \sqrt{ N_0^i N_0^j}} 
G_0^{ij} \quad {\rm if} \ i \ne j \nonumber \\
\Delta_{ii} &=& {2 \rho \over \rho_i^2} {1 \over N_0^i} 
(1 + G_0^{ii} ) \nonumber \\
G_0^{ij} &=& \sqrt{N_0^i N_0^j} g_0^{ij} \quad, \qquad  
N_0^i= {m_i^* k_{Fi} \over \pi^2 \hbar^2}
\label{eq:Dij-gij}
\eeq
A criterion for the appearance of the ferromagnetic phase is that 
the determinant of the inverse susceptibility matrix vanishes
\beq
Det \left( \matrix{1/\chi_{nn} & 1/\chi_{np} & 1/\chi_{n\Lambda} \cr
                   1/\chi_{pn} & 1/\chi_{pp} & 1/\chi_{p\Lambda} \cr
    1/\chi_{\Lambda n} & 1/\chi_{\Lambda p} & 1/\chi_{\Lambda\Lambda} \cr}
\right) =0
\label{eq:suscept}
\eeq
and in terms of the Landau parameters:
\beq
Det \left( \matrix{ (1 + G_0^{nn}) & G_0^{np} & G_0^{n\Lambda} \cr
                    G_0^{pn} & (1 + G_0^{pp}) & G_0^{p\Lambda} \cr
    G_0^{\Lambda n} & G_0^{p\Lambda} & (1 + G_0^{\Lambda\Lambda}) \cr} 
\right) =0
\label{eq:ferro}
\eeq
It can be shown that this quantity appears in the denominator of the 
static axial response (see \cite{M04b}).

\subsection{Dependence of the criterion on parameter sets}

A few NN forces (MSk7,SkX,SkS3) had to be discarded as not convenient 
for obtaining the equation of state away from  saturation, some more 
for not permitting the formation of hyperons below 5 $n_{\rm sat}$,
others again for undergoing a transition to pure neutron matter
before the threshold for hyperon formation was reached (SIII, SI', 
SIII', Skz0--Skz4).

The remaining ones all display a transition to the ferromagnetic 
state at some critical density $n_{\rm ferro}^{\beta-npe}$ when no 
hyperons are present, with the exception of the SV force. Usually the 
ferromagnetic transition occurs earlier in pure neutron matter (PNM) 
than in symmetric nuclear matter (SNM), and at an intermediate density 
for $npe$ matter in $\beta$ equilibrium. One exception is the case of 
the Lyon Skyrme forces where the reverse situation appears to occur.
A closer look nevertheless reveals that only traces of protons are 
enough to drastically lower the critical density in very neutron rich
matter so that the usual pattern is in fact recovered.

Some forces (SkO, SkO', SkI1, SkI4, SkT4, SkT5) again need to be 
discarded since the ferromagnetic transition occurs below $n_{\rm sat}$ 
in symmetric nuclear matter, so that the nuclei would in fact be unstable 
with respect to spin fluctuations. This leaves us with forces for which 
the $n_{\rm ferro}^{\beta-npe}$ occurs in the range $n_{\rm sat}$ 
-- $3\ n_{\rm sat}$, and the threshold for $\Lambda$ hyperon
formation in the range $1.7\ n_{\rm sat}$ -- $4\ n_{\rm sat}$. 
A key point is now whether $n_{\rm ferro}^{\beta-npe}$ is greater or 
lower than $n_{\rm thr}^{\Lambda}$. While the criterion for the 
onset of the ferromagnetic instability Eq. (\ref{eq:ferro})
decreases with increasing density in $npe$ matter in $\beta$ 
equilibrium, we noted that it always tends to increase again when
enough hyperons are present. This then rejects the critical density
$n_{\rm ferro}^{\beta-npe\Lambda}$ for the onset of ferromagnetism 
in $np\Lambda e$ matter in $\beta$ equilibrium to higher densities. 
In a few cases the instability even disappears altogether. For the 
majority of NN parameter sets, the ferromagnetic transition occurs 
before the threshold for hyperon formation. In some cases (SLy4, 
SLy7, SGI) we have $n_{\rm thr}^{\Lambda} \sim < n_{\rm ferro}^{\beta-npe}$,
and the ferromagnetic transition is only delayed by a tiny amount. 

\vskip 0.2cm

We finally selected four NN forces (SLy10, SkI3, SkI5, SV) as relevant 
for our purpose, namely to study neutron star matter with an hyperonic 
component before the ferromagnetic transition sets in. In Figure 1 (a--c) 
we have represented the criterion (\ref{eq:ferro}) for these parametrizations 
of the NN Skyrme force and various choices of the N$\Lambda$ and 
$\Lambda\Lambda$ forces. The modification of the slope at the
hyperon threshold around $2\ n_{\rm sat}$ is clearly visible on
Fig. 1-a.

\begin{figure}[htb]
\mbox{%
\parbox{8cm}{\epsfig{file=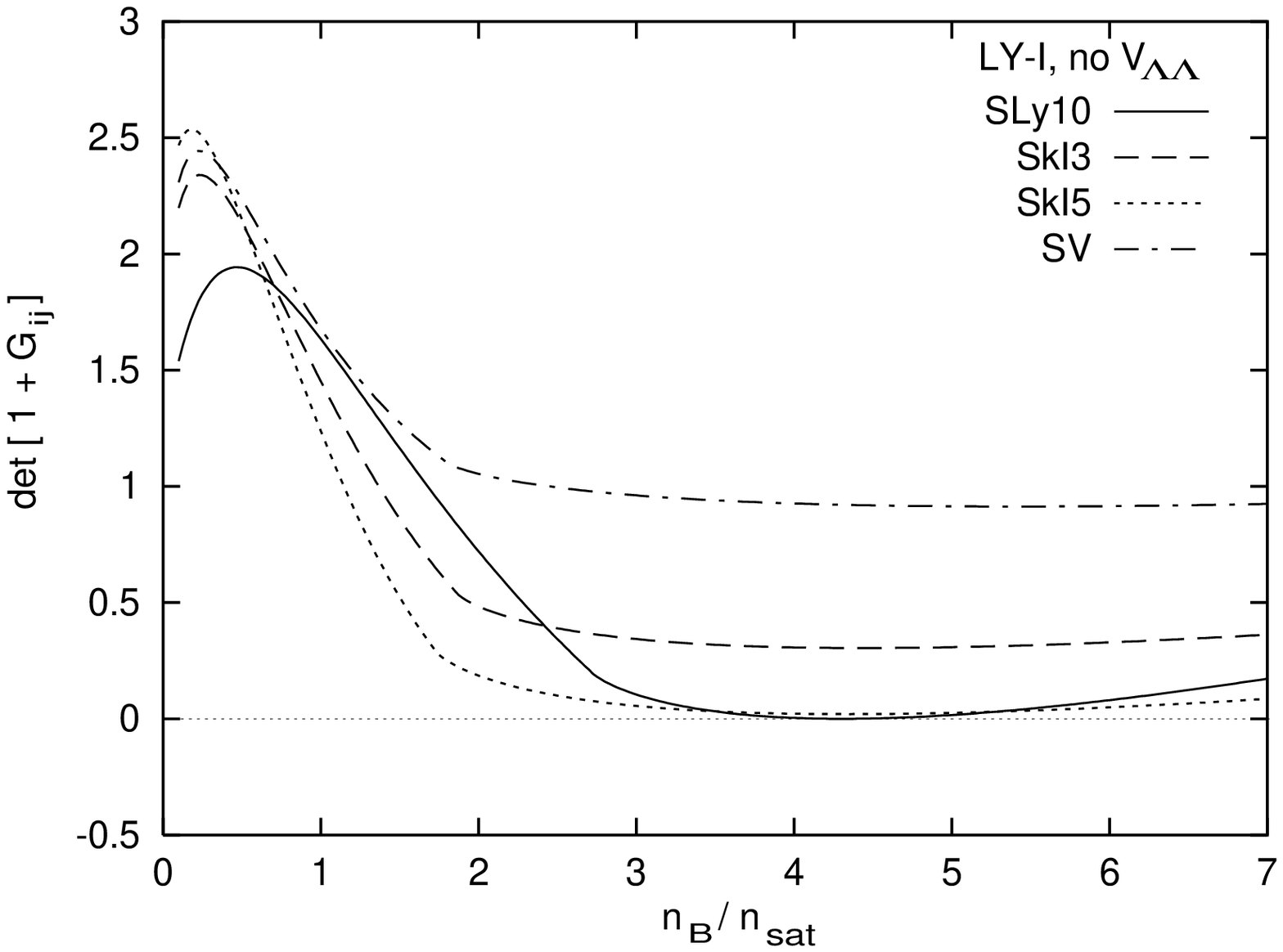,width=8cm}}
\parbox{8cm}{\epsfig{file=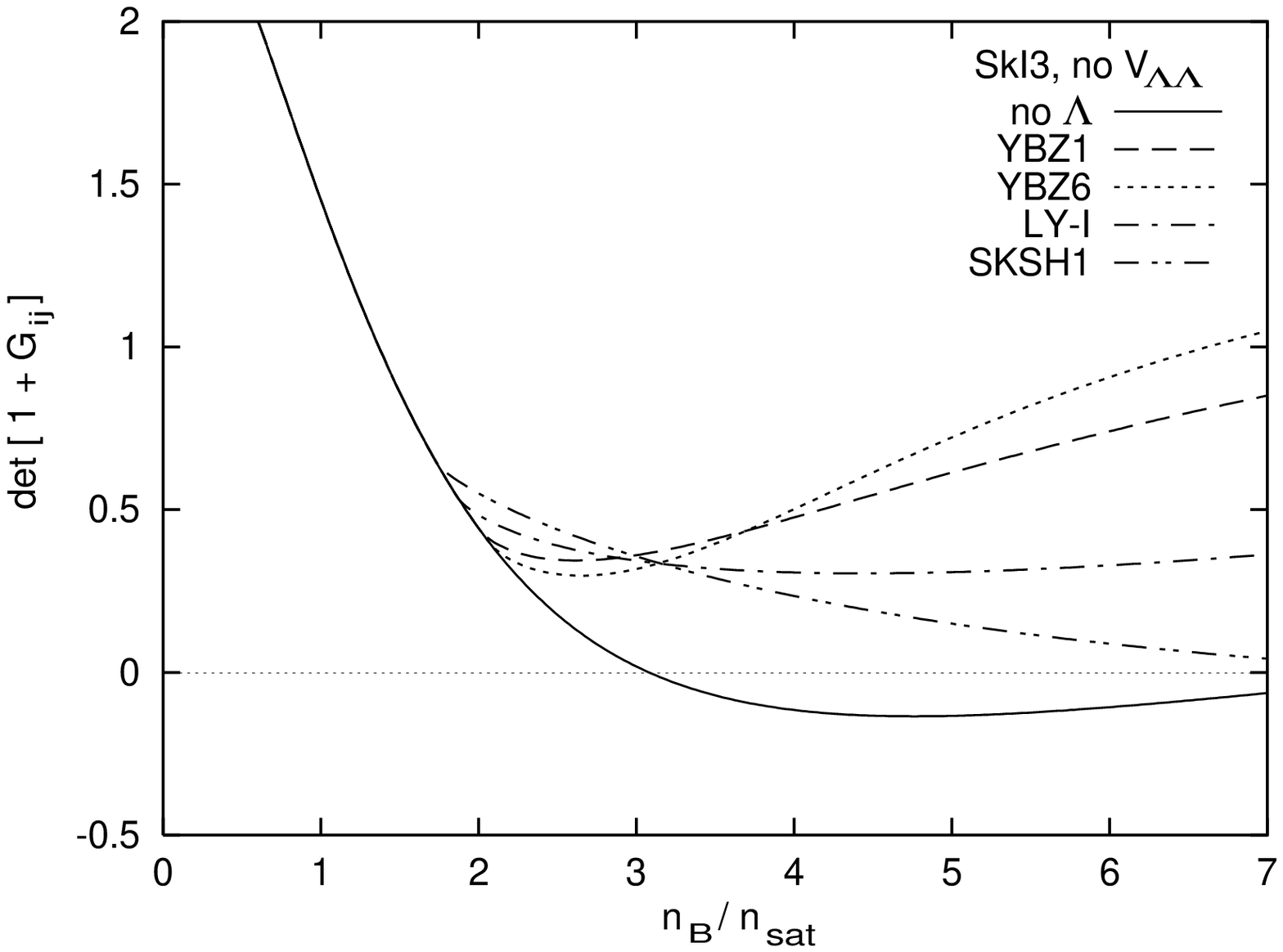,width=8cm}}
}
\vskip 0.2cm
\mbox{%
\parbox{8cm}{\epsfig{file=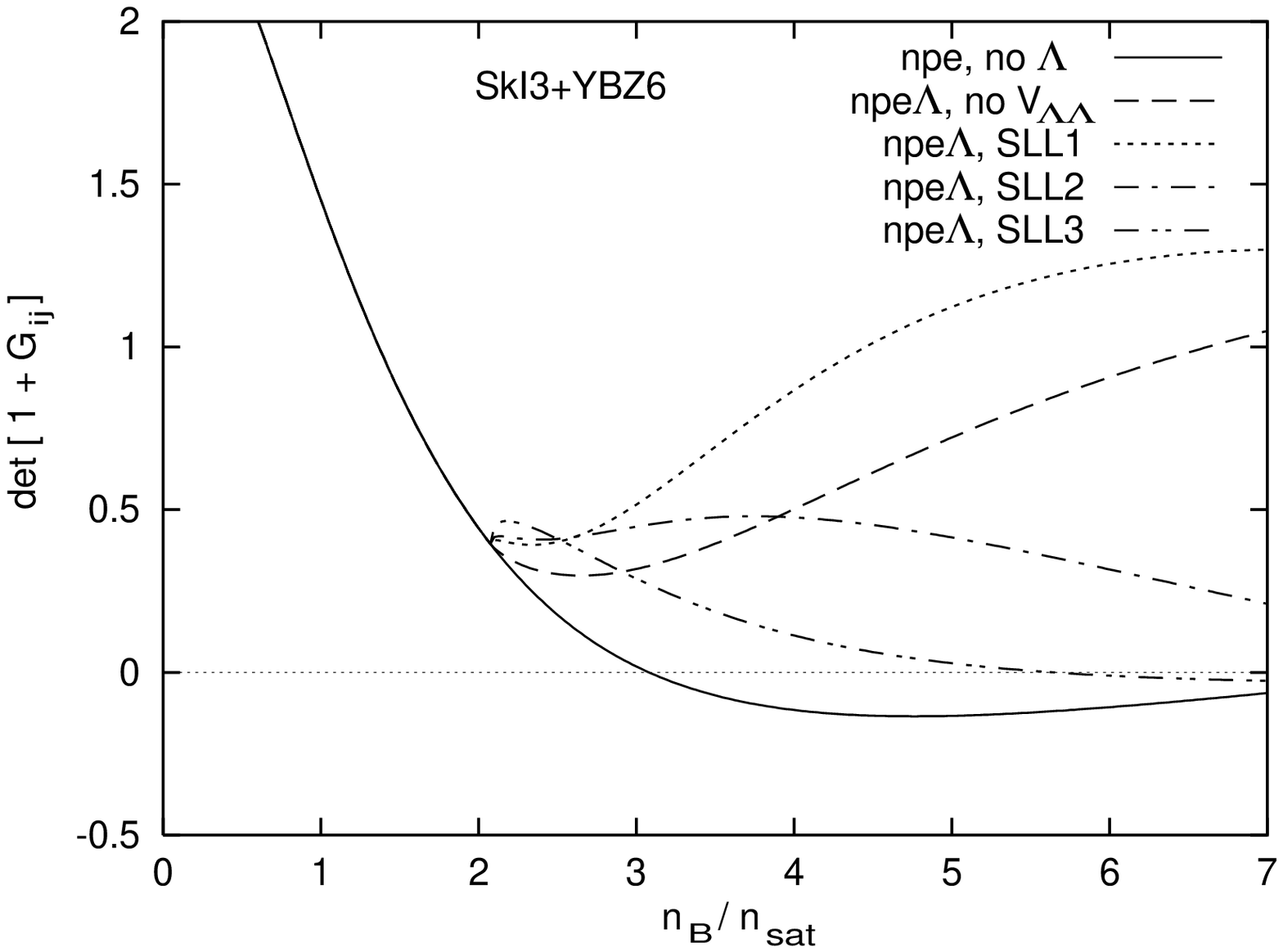,width=8cm}}
\parbox{1cm}{\phantom{aaaaa}}
\parbox{7cm}{{\bf Fig. 1} -- Criterion for the transition to 
a ferromagnetic state.}
}
\end{figure} 

The SV NN interaction is atypical as it has no $t_3$ term. This 
parametrization is especially interesting since, among all tested 
NN interactions, it was the only one which does not give a ferromagnetic 
instability in $npe$ matter in $\beta$ equilibrium. This parametrization 
is rather old (1975) but yields an acceptable description of the 
properties of nuclear matter and nuclei and it complies with all the 
conditions necessary for describing a viable neutron star. The SLy10,
SkI3 and SkI5 are modern forces. The SLy10 set was designed to reproduce 
features of pure neutron matter as obtained from the variational 
calculations of Wiringa {\it et al.} in view of its application to 
neutron stars. The SkI3 and SkI5 sets were designed in order to improve
isotope shifts and incorporate the dipole sum rule enhancement factor
$\kappa=0.25$. Let us notice that the SkI3, SkI5 and SV allow for 
larger proton fractions than the SLy10, in particular we can see 
from table IV that the criterion for opening of the direct URCA process 
with nucleons is reached before the theshold for hyperon production.

\vskip 0.2cm

The behaviour of the criterion for ferromagnetic instability is qualitatively 
very similar for all choices of the N$\Lambda$ interaction. Forces yielding 
smaller $\rho_\Lambda(thr)$ and softer eos (see \S \ref{S:eos-MvsR}) also
tend to be more efficient in lifting the ferromagnetic criterion
${\rm det}[1+G_{ij}]$ above the critical zero axis. This is especially the
case for the $YBZ4$ parameter set which permitted to avoid the pole also 
in the SLy4 and SLy7 parametrizations of the NN force. However the YBZ4
parameter set was rejected by Lanskoy on the ground that it gives 
overbinding of the $\Lambda$ in hypernuclei so that we will not consider 
it further. Despite this overall similarity the actual presence and 
position of the pole is sensitive to the choice of the N$\Lambda$ force. This 
happens because the ${\rm det}[1+G_{ij}]$ has already decreased considerably 
prior to $\rho_\Lambda(thr)$ and the turnover due to the contribution 
of the $\Lambda$ therefore must take place in the vicinity of the zero axis.

\vskip 0.2cm

We can also see on this figure that the $\Lambda-\Lambda$ interaction 
plays an important role in determining $n_{\rm ferro}^{\beta-npe\Lambda}$.
The set SLL3 (which also gives a stiffer eos, see next section) is less 
efficient in removing the pole while the SLL1 set which gives the softest
eos is also the most efficient in preventing the onset of ferromagnetism.
The $\Lambda$-$\Lambda$ force of Lanskoy \cite{L98} is very schematic 
(it is parametrized by $\lambda_0$ and $\lambda_1$ only, see the Appendix);
moreover it was adjusted to the older value $\Delta B_{\Lambda\Lambda}
= -4.8$ MeV instead of the value $\Delta B_{\Lambda\Lambda}
\simeq -1$ MeV recently extracted from the ``Nagara event''. The
case where a vanishing $\Lambda$-$\Lambda$ interaction is assumed may 
in fact be closer to the true situation.  A better knowledge of the 
$\Lambda$-$\Lambda$ interaction at high density is needed.

\vskip 0.2cm

The values of density and proton content at the threshold for hyperon
formation as well as the critical values for the ferromagnetic transition 
for various models are gathered in tables IV a--d. All densities are 
quoted in units of the saturation density of the corresponding model. 
The mention ``grazing'' means that ${\rm det}[1-G_{ij}]$, while not
actually crossing the zero axis, comes so near it that for all practical 
purposes the static axial response function will behave as if a pole 
were present.

\begin{center}
\begin{flushleft}
$\underline{\mbox{{\sf a) }\rm NN force=SLy10}}$ 
\smallskip \par [threshold for ferromagnetism in PNM: 3.837 $n_{\rm sat}$, 
in SNM: 3.821 $n_{\rm sat}$, in $npe$ matter in $\beta$-equilibrium: 
3.055 $n_{\rm sat}$] 
\end{flushleft}
\vskip 0.2cm
\begin{tabular}{|l||c|c||c|c|c|c|}
\hline
 N$\Lambda$    & $n_{\rm thr}$ & $Y_p$ & no $\Lambda\Lambda$ 
& SLL1 & SLL2 & SLL3 \\
\hline
LY-I   & 2.719 & 0.041 & grazing [4.3] & no pole & 6.183 & 3.898 \\
LY-II  & 2.868 & 0.040 & 3.160         & 3.212   & 3.236 & 3.328 \\
LY-IV  & 2.764 & 0.041 & 3.657         & no pole & 6.480 & 3.818 \\
YBZ5   & 3.319 & 0.037 &  --           &  --     &  --   &  --   \\
YBZ6   & 5.101 & 0.029 &  --           &  --     &  --   &  --   \\
SKSH1  & 2.294 & 0.044 & 4.577         &  5.297 & 4.729 & 4.153 \\
\hline
\end{tabular}
\end{center}
\vskip 1cm

\begin{center}
\begin{flushleft}
$\underline{\mbox{{\sf b) }\rm NN force=SkI3}}$ 
\smallskip \par [threshold for ferromagnetism in PNM: 2.298 $n_{\rm sat}$, 
in SNM: 5.73 $n_{\rm sat}$, in $npe$ matter in $\beta$-equilibrium: 
3.078 $n_{\rm sat}$] 
\end{flushleft}
\vskip 0.2cm
\begin{tabular}{|l||c|c||c|c|c|c|}
\hline
 N$\Lambda$    & $n_{\rm thr}$ & $Y_p$ & no $\Lambda\Lambda$ 
& SLL1 & SLL2 & SLL3 \\
\hline
LY-I   & 1.864 & 0.126 & no pole & no pole & 9.234 & 4.663 \\ 
LY-IV  & 1.873 & 0.127 & no pole & no pole & 8.900 & 4.548 \\
YBZ1   & 2.015 & 0.139 & no pole & no pole & 8.753 & 4.893 \\
YBZ5   & 1.942 & 0.133 & no pole & no pole & 7.550 & 4.080 \\
YBZ6   & 2.076 & 0.144 & no pole & no pole & 9.194 & 5.633 \\
SKSH1  & 1.739 & 0.116 & 8.225   & 9.495   & 6.765 & 4.425 \\
SKSH2  & 1.681 & 0.111 & 18.875  & 15.250  & 8.253 & 4.212 \\
\hline
\end{tabular}
\end{center}
\vskip 0.2cm

\begin{center}
\begin{flushleft}
$\underline{\mbox{{\sf c) }\rm NN force=SkI5}}$ 
\smallskip \par [threshold for ferromagnetism in PNM: 1.772 $n_{\rm sat}$, 
in SNM: 2.659 $n_{\rm sat}$, in $npe$ matter in $\beta$-equilibrium: 
2.140 $n_{\rm sat}$] 
\end{flushleft}
\vskip 0.2cm
\begin{tabular}{|l||c|c||c|c|c|c|}
\hline
 N$\Lambda$    & $n_{\rm thr}$ & $Y_p$ & no $\Lambda\Lambda$ 
& SLL1 & SLL2 & SLL3 \\
\hline
LY-I   & 1.727 & 0.150 & grazing [4.34] & no pole & 4.707 & 3.166 \\
LY-IV  & 1.734 & 0.150 & grazing [4.14] & no pole & 4.627 & 3.095 \\ 
YBZ1   & 1.838 & 0.162 & no pole        & no pole & 8.656 & 3.622 \\
YBZ5   & 1.785 & 0.156 & no pole        & no pole & 5.404 & 3.239 \\
YBZ6   & 1.882 & 0.167 & no pole        & no pole & 9.118 & 3.950 \\
SKSH1  & 1.629 & 0.138 & 3.337          & 3.914   & 3.541 & 3.103 \\
\hline
\end{tabular}
\end{center}
\vskip 0.2cm

\begin{center}
\begin{flushleft}
$\underline{\mbox{{\sf d) }\rm NN force=SV}}$ 
\smallskip \par [threshold for ferromagnetism in PNM: 4.850 $n_{\rm sat}$, 
in SNM: no pole, in $npe$ matter in $\beta$-equilibrium: no pole] 
\end{flushleft}
\vskip 0.2cm
\begin{tabular}{|l||c|c||c|c|c|c|}
\hline
 N$\Lambda$    & $n_{\rm thr}$ & $Y_p$ & no $\Lambda\Lambda$ 
& SLL1 & SLL2 & SLL3 \\
\hline
LY-I   & 1.793 & 0.125 & no pole & no pole & 9.330  & 4.576 \\
LY-IV  & 1.800 & 0.126 & no pole & no pole & 9.143  & 4.533 \\ 
YBZ1   & 1.906 & 0.133 & no pole & no pole & 8.649  & 4.635 \\
YBZ3   & 1.672 & 0.116 & no pole & no pole & 11.700 & 5.050 \\
YBZ5   & 1.851 & 0.129 & no pole & no pole & 7.521  & 3.956 \\
YBZ6   & 1.951 & 0.136 & no pole & no pole & 9.093  & 5.200 \\
SKSH1  & 1.691 & 0.118 & no pole & no pole & 8.920  & 4.250 \\
SKSH2  & 1.635 & 0.114 & no pole & no pole & 8.253  & 4.068 \\
\hline
\end{tabular}
\end{center}
\vskip 0.2cm

\begin{center}
{\bf Tables IV a--d}: thresholds for ferromagnetism in $np\Lambda e$ matter in 
$\beta$ equilibrium 
\end{center}

\section{Equation of state and neutron star structure}
\label{S:eos-MvsR}

\subsection{Equation of state, effective masses and the hyperon fraction}

For neutron star matter in beta equilibrium without hyperons, the equations 
of state are in order of decreasing stiffness parametrized by the NN 
interaction SV $>$ SkI3, SkI5 $>$ SLy10. The parametrizations SkI3 and SkI5 
yield nearly indistinguishable results, so that SkI5 will not considered 
further in the remainder of this work. When the hyperons are taken into 
account in the calculation of the $\beta$ equilibrium, the equation of 
state softens as expected. For a given NN interaction, we have in order 
of decreasing stiffness YBZ6 $>$ YBZ1 $>$ LY-I, LY-IV $>$ YBZ3 $>$ SKSH2 
$>$ SKSH1. 
Finally, when varying the $\Lambda\Lambda$ interaction for given NN and 
N$\Lambda$ forces, we obtain SLL3 $>$ SLL2 $>$ SLL1. If the $\Lambda\Lambda$ 
interaction is set to zero, the equation of state is somewhat stiffer than 
SLL2 at $n_B < 5\ n_{\rm sat}$ and much softer above $5\ n_{\rm sat}$.
We note the sizable effect of the $\Lambda$-$\Lambda$ interaction on the 
equation of state, as well as in the previous section on the value of
the Landau parameters.

\begin{figure}[htb]
\mbox{%
\parbox{8cm}{\epsfig{file=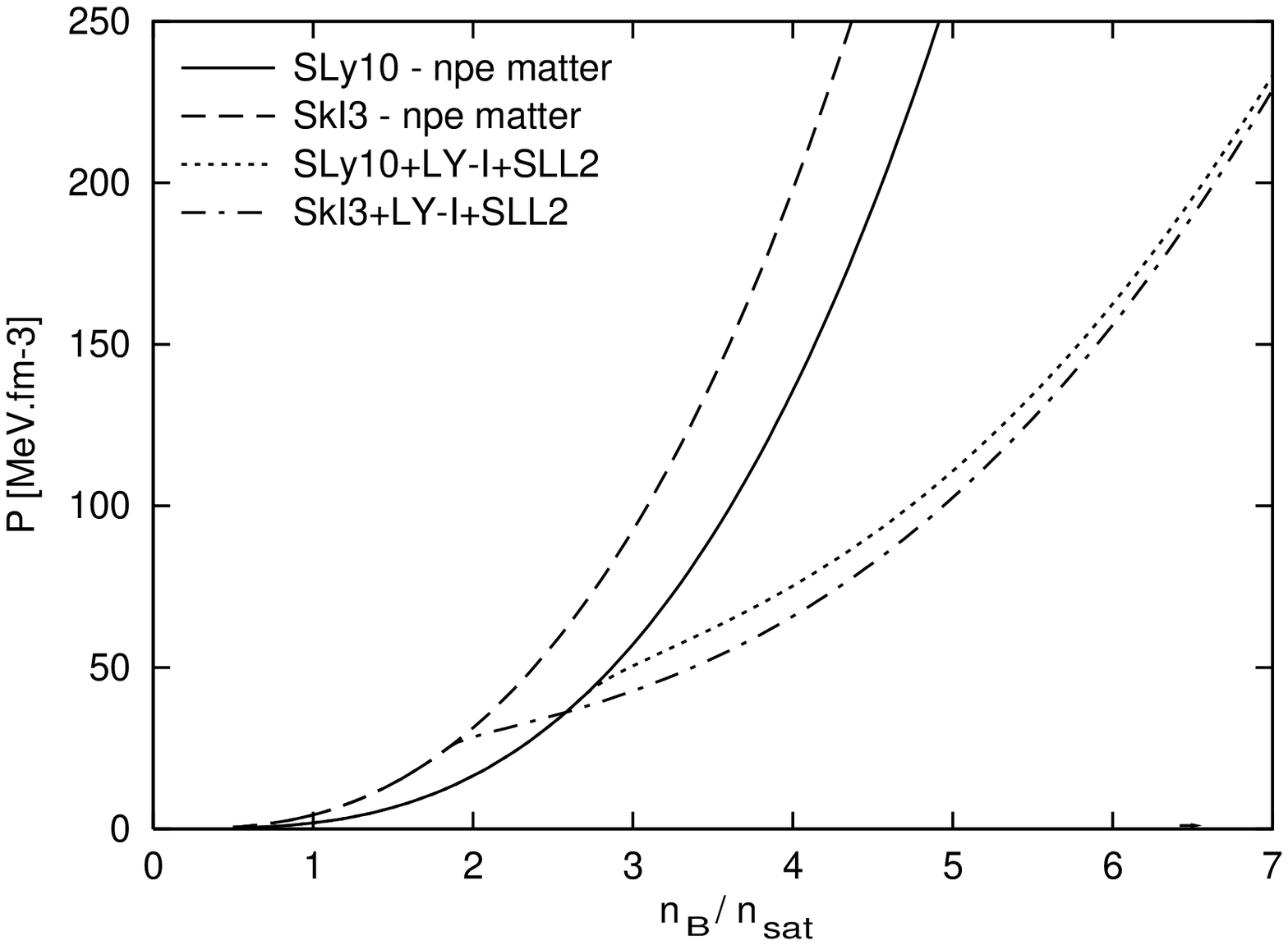,width=8cm}}
\parbox{8cm}{\epsfig{file=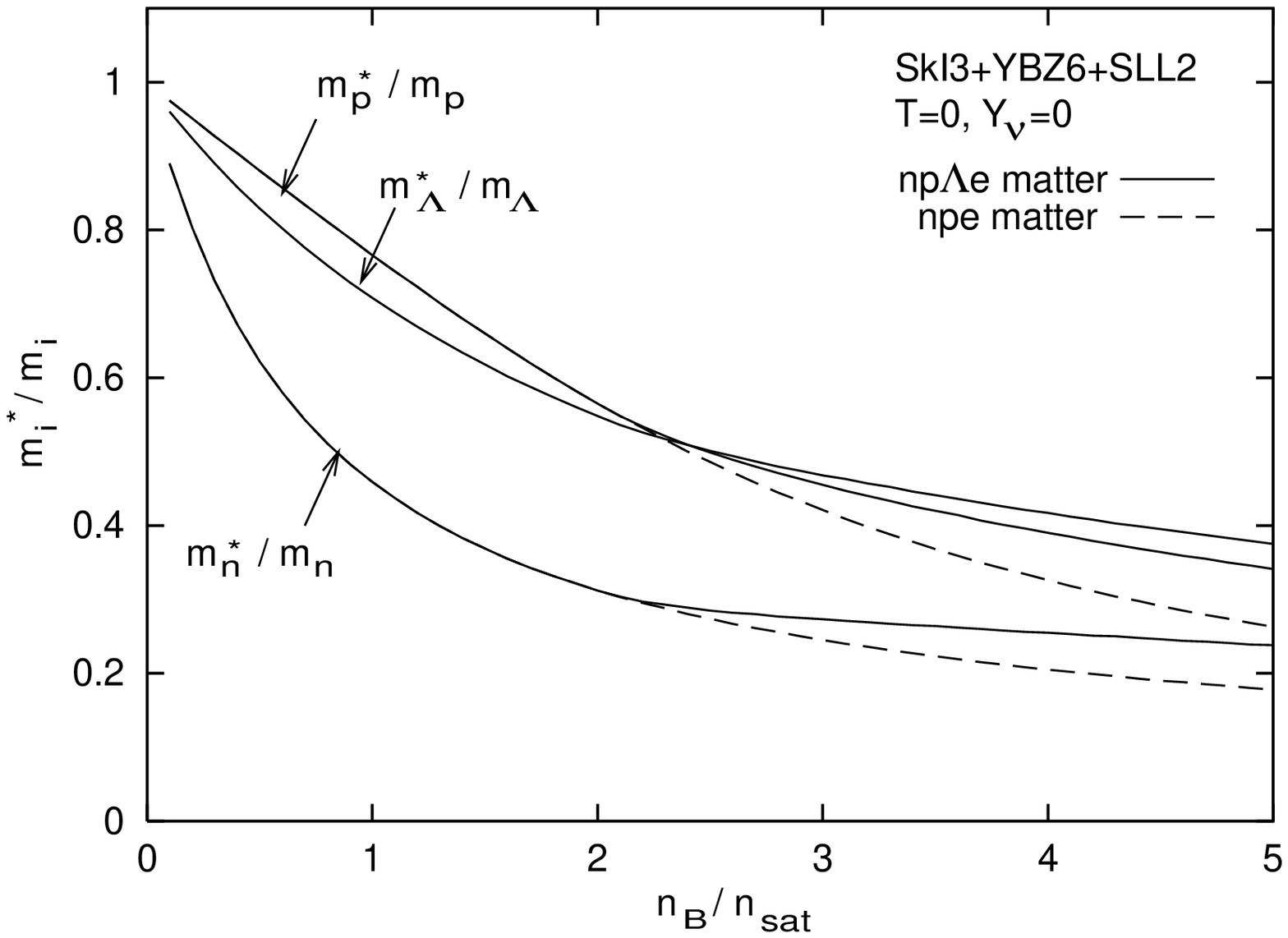,width=8cm}}
}
\vskip 0.2cm
\mbox{%
\hskip 0.5cm
\parbox{7cm}{{\bf Fig. 2} -- Typical results for the equation of state}
\parbox{1.5cm}{\phantom{aaa}}
\parbox{7cm}{{\bf Fig. 3} -- Modification of the effective masses by
the presence of hyperons.}
}
\end{figure} 
\vskip 0.2cm

The equation of state is shown on Fig. 2 for two of our preferred NN forces,
and with the same choice LY-I for the N$\Lambda$ interaction and SLL2 for 
the $\Lambda\Lambda$ interaction. Even though SkI3 was stiffer that SLy10
without hyperons, the combination SkI3+LYI+SLL2 gives rise to a higher hyperon
content at a given density (see also Fig. 4) than SLy10+LYI+SLL2, so that
it is also subject to more softening. As a consequence, the eos with hyperons
are very similar.

Figure 3 illustrates the behavior of the  effective masses in the case of 
the parameter set SkI3 +YBZ6+SLL2. The effective mass of the neutrons are 
lower than that of protons in neutron rich matter, a feature generally 
encountered in Skyrme models. The presence of hyperons cause both the 
neutron and proton mass to decrease less rapidly at high density.

\begin{figure}[htb]
\mbox{%
\parbox{8cm}{\epsfig{file=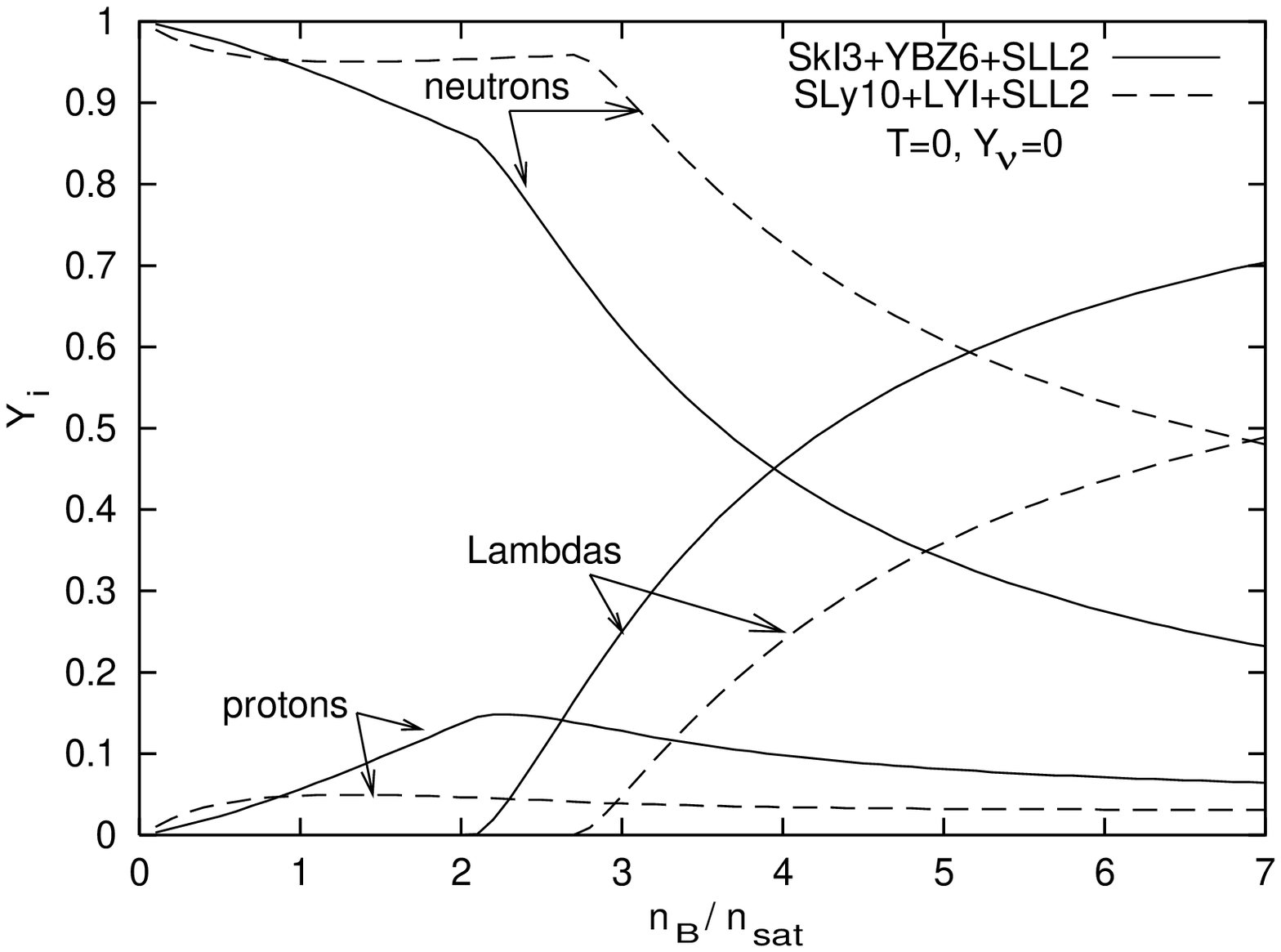,width=8cm}}
\parbox{8cm}{\epsfig{file=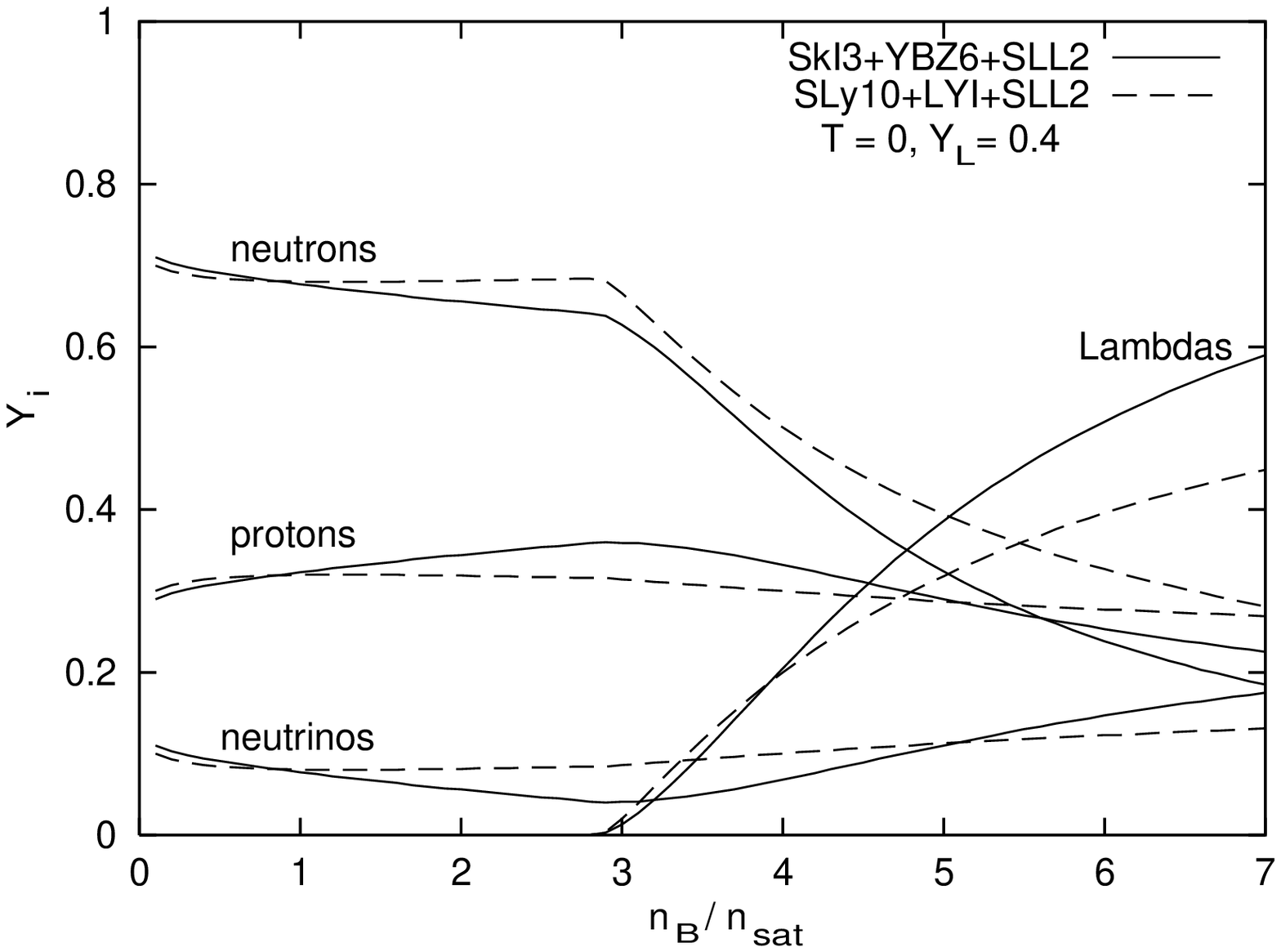,width=8cm}}
}
\vskip 0.2cm
\mbox{%
\parbox{16cm}{{\bf Fig. 4} -- Particle fractions from the SkI3+YBZ6+SLL2 and
  SLy10+LYI+SLL2 parametrizations. Left: in neutrino-free matter, right:
with trapped neutrinos and lepton fraction $Y_L=0.4$}
}
\end{figure} 
\vskip 0.4cm

Figure 4 shows the particle fractions $Y_i = n_i/n_{B}$ as a 
function of total baryonic density $n_B$ for the  SkI3+YBZ6+SLL2 and
SLy10+LYI+SLL2 parameter sets. It can be seen on the left panel 
that the hyperons appear at higher density and are less numerous
for a given density with SLy10+LYI+SLL2 than with SkI3+YBZ6+SLL2.
The right panel shows the chemical composition of matter with
a non vanishing number of trapped neutrinos at zero temperature. 
The case with trapped neutrinos and finite temperature relevant for
protoneutron stars will be discussed further in \S \ref{S:finiteT}

\subsection{Solution of the Tolman-Oppenheimer-Volkoff equation}

Our selected parameter sets still have to pass the test of causality
in the density range of interest, and whether they can support neutron
stars with maximum masses larger than the observed value 1.4 $M_\odot$.

The properties of neutron stars formed of $npe$ matter in $\beta$ equilibium 
were calculated by  Rikovska Stone {\it et al} \cite{RSMKSS03} for a 
large number of Skyrme parametrizations. The four NN interactions that 
we singled out in section \S \ref{S:ferro} all belong to the subset of 
interactions selected by these authors as giving viable neutron stars, 
with maximum masses of the order 2 $M_\odot$ to 2.4 $M_\odot$. The central 
density reached in stars with the maximum mass is slightly larger than 
the density at which the velocity of sound reaches the velocity of light 
for the SLy10 and SV parametrizations whereas the equation of state 
obtained with SkI3 and SkI5 always remains causal. In any case, stars 
with the fiducial mass 1.4 $M_\odot$ always fulfill $n_c (1.4\ M_\odot)
< n_B(c_s^2=1)$.

When hyperons are added, the equation of state being softer, the limit
$c_s^2=1$ is reached for larger densities. It is pushed to $\sim$ 8 --
11 times saturation density for the SLL3 parametrization of the 
$\Lambda$-$\Lambda$ interaction, to $\sim$ 13 -- 16 $n_{\rm sat}$ 
for the SLL2 parametrization and is larger than $20\ n_{\rm sat}$ 
for the SLL1 set. The case where a vanishing $\Lambda$-$\Lambda$ 
interaction is assumed always remains causal. 
In any case, such densities are beyond the range of validity of the model.
As a consequence, although the softening of the eos with hyperons gives 
rise to higher compression rates and larger central densities in neutron 
stars than in stars made of $npe$ matter only, the criterion $c_s^2<1$ 
is always fulfilled up to the central density of the most massive stars
in our neutron star models with hyperons, except a few instances involving 
the SLL3 parameter set.

\vskip 0.2cm

We have solved the Tolman-Oppenheimer-Volkoff equation to obtain the 
(non-rotating) neutron star masss-radius relation. The equation of state 
is matched at lower densities with that of Negele and Vautherin \cite{NV73} 
for $\rho\in [0.001-0.08]\ {\rm fm}^{-3}$ and with the Baym-Pethick-Sutherland
(BPS) \cite{BPS71} equation of state for $\rho < 0.001\ {\rm fm}^{-3}$.
The results\footnote{The parameters we obtained for the stars composed of 
$npe$ matter may differ very slightly from those quoted by Rikovska Stone 
{\it et al.}, presumably due to a different matching at low density or the
neglect of the muons.} are summarized in Table V a-d and a sample of 
mass-radius curves is shown in Fig. 5 and 6. For each combination of 
the NN, N$\Lambda$ and $\Lambda\Lambda$ interaction we give the density 
at which the speed of sound becomes superluminal, the central density
and radius of a $1.4\ M_\odot$ in case it can be formed, and the central 
density and radius of the star with maximum mass $M_{\rm max}$. 

\begin{figure}[htb]
\mbox{%
\parbox{10cm}{\epsfig{file=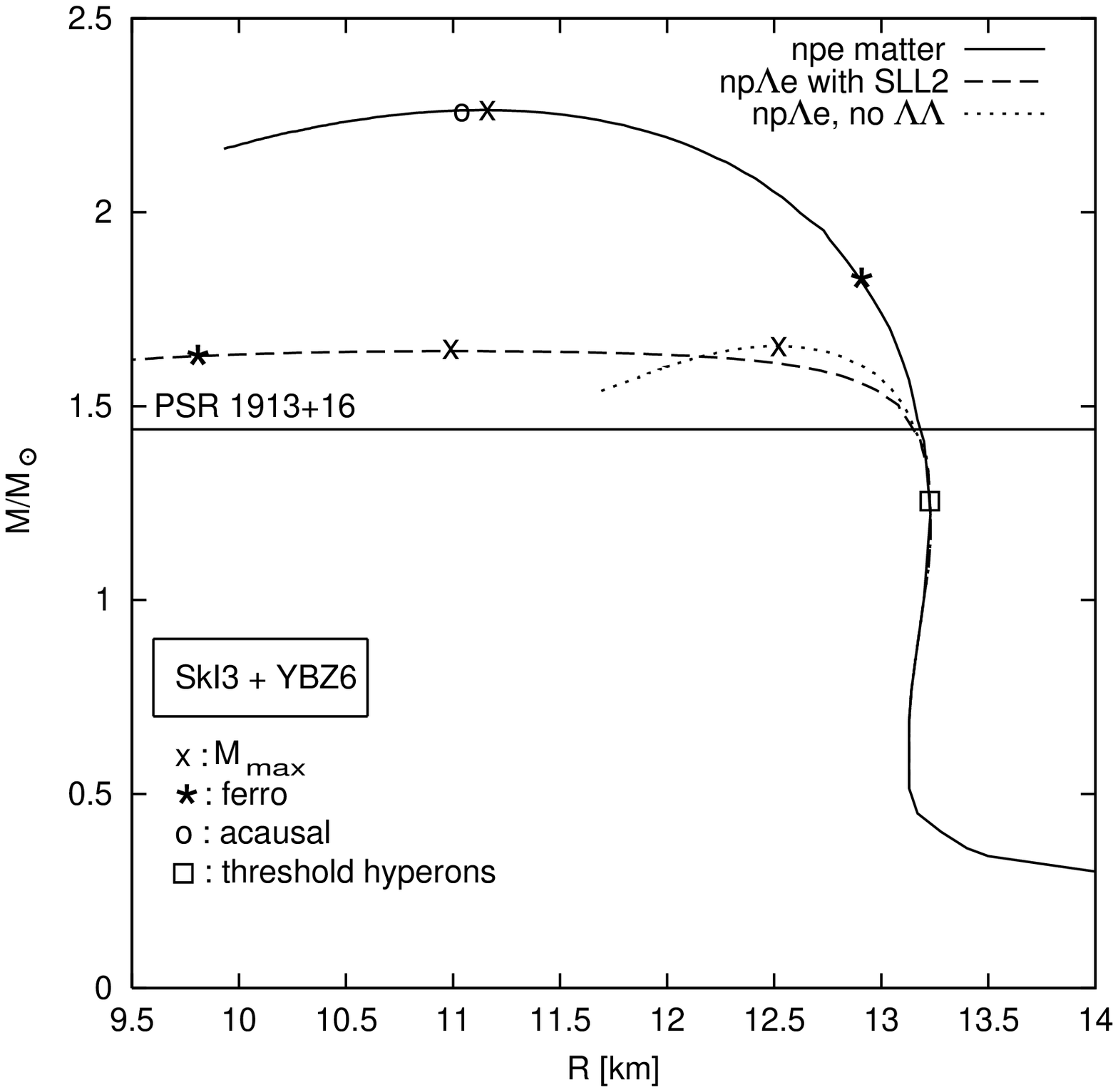,width=10cm}}
\parbox{7cm}{{\bf Fig. 5} -- Neutron star mass-radius relation with
the Skyrme SkI3+YBZ6 interaction, with and without hyperons.}
}
\end{figure} 

As a rule of thumb it is known that the stiffer the equation 
of state is, the larger  will be the maximum mass of the neutron star. 
Our results for $M_{\rm max}$ follow accordingly to the classification
in stiffness given at the beginning of this section. Thus the SKSH1
choice generally yields $M_{\rm max}<1.4\ M_\odot$ whereas the YBZ6
always pass this test successfully. We see again that the choice of 
the $\Lambda$-$\Lambda$ interaction is crucial to determine whether or 
not the star is able to reach to 1.4 $M_\odot$ line. When the
$\Lambda$-$\Lambda$ interaction is switched off or for the SLL1 choice,
the equation of state is generally too soft to support a 1.4 neutron star.
On the other hand the SLL3 always succeed in producing $M_{\rm max}>
1.4\ M_\odot$. The SLL2 choice barely makes it to $1.4\ M_\odot$ for
intermediate choices of the N$\Lambda$ interaction. This results in a
plateau feature, where tiny additions of mass effect a large reduction 
of radius without actually reaching the instability, the $M=1.4\ M_\odot$ 
point being finally reached at unprobably high densities $\rho > 
10\ n_{\rm sat}$ much beyond the validity of the model. 
A stiff equation of state on the other hand is not convenient since 
it favours ferromagnetism to appear earlier. The latter condition 
usually rules out the SLL3 choice.

For the SLy10 choice of the NN interaction ferromagnetism always sets 
in before the central density of a $1.4\ M_\odot$ star is reached. 
In particular, it must be reminded at this point that ferromagnetism 
is reached before the threshold of production of hyperons for the 
choices YBZ1, YBZ5, YBZ6. For the SkI3 choice, the criterion 
$n_c (1.4\ M_\odot)< n_{\rm ferro}$ is satisfied for the sets 
(SkI3 + LY-IV + SLL3), (SkI3 + YBZ5 + SLL3) and (SkI3 + YBZ1 or 
YBZ6 + any choice of $\Lambda\Lambda$ ). The star with the maximum 
mass also fulfills all requirements if it is described by 
(SkI3 + YBZ6 + no $\Lambda\Lambda$ ) or  (SkI3 + YBZ6 + SLL2). 
The results obtained with the choice SkI5 are very 
similar. The choice SV represents an intermediate situation. The 
$1.4\ M_\odot$ star fulfills both criteria $n_c (1.4\ M_\odot)< 
n_B(c_s^2=1)$ and $n_c (1.4\ M_\odot)< n_{\rm ferro}$ in the cases 
(SV + LY-I + SLL3), (SV + LY-IV + SLL2), (SV + YBZ1 + SLL2/3), 
(SV + YBZ6 + any choice of $\Lambda\Lambda$). Moreover the star 
with the maximum mass also fulfills both criteria for the choices 
(SV + YBZ6 + no $\Lambda\Lambda$) and (SV + YBZ6 + SLL2).

\newpage

\begin{center}
\begin{flushleft}
$\underline{\mbox{{\sf a) }\rm NN force=SLy10}}$ 
\smallskip \par [in $npe$ matter in $\beta$-equilibrium: 
$n_B (c_s^2=1)=7.308\ n_{\rm sat}$, $n_c (1.4\ M_\odot) 
=3.60\ n_{\rm sat}$, $R (1.4\ M_\odot)$=11.05 km, 
$n_{\rm max}=7.69\ n_{\rm sat}$, $M_{\rm max}=1.99\ M_\odot$, 
$R(M_{\rm max})=9.52\ $km]  
\end{flushleft}
\vskip 0.2cm
\begin{tabular}{|c|c||c||c|c||c|c|c|}
\hline
N$\Lambda$  &  $\Lambda\Lambda$ & $n_B(c_s^2=1)$ & $n_c (1.4\ M_\odot)$ 
& $R (1.4\ M_\odot)$ & $n_{\rm max}$ & $M_{\rm max}$ & $R(M_{\rm max})$ \\
\hline
\hline 
LY-I  & no $\Lambda\Lambda$ & causal &   -  &   -   &  6.94 & 1.317 & 10.36 \\
LY-I  & SLL1  & $>20$  &   -  &   -   &  5.99 & 1.210 & 10.70 \\
LY-I  & SLL2  & 16.032 & 9.38 &  9.0  & 12.38 & 1.425 & 8.11 \\
LY-I  & SLL3  & 11.036 & 5.25 & 10.42 & 10.23 & 1.658 & 8.55 \\
\hline
LY-IV & no $\Lambda\Lambda$ & causal &  -   &  -   & 6.82  & 1.338 & 10.43 \\
LY-IV & SLL2  & 15.880 & 8.49 & 9.35 & 12.07 & 1.437 & 8.21  \\
\hline
SKSH1 & no $\Lambda\Lambda$ & $>20$  &   -   &   -   & 5.09  & 0.875 & 10.8 \\
SKSH1 & SLL2  & $>20$  &   -   &   -   & 20.6  & 1.159 & 6.19 \\
SKSH1 & SLL3  & 13.210 & 10.25 &  8.15 & 13.65 & 1.453 & 7.40 \\
\hline
YBZ5  & no $\Lambda\Lambda$ & causal &  -  &   -   & 5.51  & 1.377 & 10.98 \\
YBZ5  & SLL1  & $>20$  &  -  &   -   & 4.94  & 1.234 & 11.11 \\
YBZ5  & SLL2  & 13.825 & 7.11 & 10.04 & 11.64 & 1.456 & 8.34 \\
YBZ5  & SLL3  & 8.440 & 4.14 & 10.97 & 9.41  & 1.789 & 8.92 \\
\hline
\end{tabular}
\end{center}
\vskip 0.2cm

\begin{center}
\begin{flushleft}
$\underline{\mbox{{\sf b) }\rm NN force=SkI3}}$ 
\smallskip \par [in $npe$ matter in $\beta$-equilibrium: 
$n_B (c_s^2=1)=6.343\ n_{\rm sat}$, $n_c (1.4\ M_\odot)=
2.27\ n_{\rm sat}$, $R (1.4\ M_\odot)$=13.21 km, $n_{\rm max}=
6.12\ n_{\rm sat}$, $M_{\rm max}=2.263\ M_\odot$, 
$R(M_{\rm max})=11.16\ $km]  
\end{flushleft}
\vskip 0.2cm
\begin{tabular}{|c|c||c||c|c||c|c|c|}
\hline
N$\Lambda$  &  $\Lambda\Lambda$ & $n_B (c_s^2=1)$ & $n_c (1.4\ M_\odot)$ 
& $R (1.4\ M_\odot)$ & $n_{\rm max}$ & $M_{\rm max}$ & $R(M_{\rm max})$ \\
\hline
\hline
LY-I  & no $\Lambda\Lambda$ & causal &  -    &   -   & 4.34  & 1.339 & 12.57 \\
LY-I  & SLL1  & $>20$  &  -    &   -   & 3.71  & 1.263 & 12.98 \\
LY-I  & SLL2  & 15.880 & 10.44 &  8.99 & 12.59 & 1.413 & 8.30 \\
LY-I  & SLL3  & 10.514 & 3.96  & 12.24 & 9.50  & 1.723 & 9.19 \\
\hline
LY-IV & no $\Lambda\Lambda$  & causal &  -   &   -   &  4.33 & 1.351 & 12.68 \\
LY-IV & SLL2  & 15.739 & 9.59 & 9.34  & 12.44 & 1.422 & 8.35 \\
LY-IV & SLL3  & 10.400 & 3.83 & 12.37 &  9.36 & 1.734 & 9.25 \\
\hline
YBZ1  & no $\Lambda\Lambda$ & causal & 2.52 & 13.20 & 4.49 & 1.534 & 12.59 \\
YBZ1  & SLL2  & 13.293 & 2.75 & 13.13 & 9.53 & 1.545 & 9.51  \\
YBZ1  & SLL3  & 7.764  & 2.68 & 13.17 & 7.93 & 1.907 & 9.85  \\
\hline
YBZ5  & SLL2 & 13.969 & 9.57 & 9.27  & 12.56 & 1.427 & 8.22 \\
YBZ5  & SLL3 & 8.269  & 3.43 & 12.69 & 8.74 & 1.822 & 9.37 \\
\hline
YBZ6  & no $\Lambda\Lambda$ & causal & 2.36 & 13.20 & 4.66 & 1.655 & 12.52 \\
YBZ6  & SLL1  & $>20$ & 2.38 & 13.19 & 4.12 & 1.553 & 12.82 \\
YBZ6  & SLL2  & 13.077 & 2.39 & 13.19 & 6.79 & 1.642 & 11.02 \\
YBZ6  & SLL3  & 7.439 & 2.41 & 13.18  & 7.43 & 1.967 & 10.16 \\
\hline
SKSH1 & no $\Lambda\Lambda$ & $>20$ &   -  &   -  & 3.32  & 1.116 & 13.02 \\
SKSH1 & SLL2  & $>20$ &   -  &   -  & 3.07  & 1.073 & 13.11 \\
SKSH1 & SLL3  & 12.46 & 7.25 & 9.91 & 12.08 & 1.544 & 8.20  \\
\hline
SKSH2 & SLL2 & 15.497 &  -   &   -   &  3.16 & 1.041 & 13.0 \\
SKSH2 & SLL3 & 9.971  & 5.81 & 10.55 & 10.83 & 1.658 & 8.46 \\
\hline
\end{tabular}
\end{center}
\vskip 0.2cm

\newpage
\begin{center}
\begin{flushleft}
$\underline{\mbox{{\sf c) }\rm NN force=SkI5}}$ 
\smallskip \par [in $npe$ matter in $\beta$-equilibrium: 
$n_B (c_s^2=1)= 6.377 \ n_{\rm sat}$, $n_c (1.4\ M_\odot)=
2.10\ n_{\rm sat}$, $R (1.4\ M_\odot)$=13.87 km, $n_{\rm max}=
6.08\ n_{\rm sat}$, $M_{\rm max}=2.273\ M_\odot$, 
$R(M_{\rm max})=11.36\ $km]  
\end{flushleft}
\vskip 0.2cm
\begin{tabular}{|c|c||c||c|c||c|c|c|}
\hline
N$\Lambda$  & $\Lambda\Lambda$ & $n_B (c_s^2=1)$ & $n_c (1.4\ M_\odot)$ 
& $R (1.4\ M_\odot)$ & $n_{\rm max}$ & $M_{\rm max}$ & $R(M_{\rm max})$ \\
\hline
\hline
LY-I  & SLL2 & 15.915 &  -   &   -   & 3.70 & 1.297 & 13.32 \\
LY-I  & SLL3 & 10.544 & 3.84 & 12.77 & 9.52 & 1.719 & 9.27 \\
\hline
YBZ6  & no $\Lambda\Lambda$ & causal & 2.20 & 13.85 & 4.44 & 1.627 & 12.96 \\
YBZ6  & SLL2  & 13.103 & 2.25 & 13.84 & 7.64 & 1.615 & 10.65 \\
YBZ6  & SLL3  & 7.470 & 2.28 & 13.85 & 7.48 & 1.958 & 10.22 \\
\hline
SKSH1 & SLL2 & $>20$ &  -  &   -   &  2.80 & 1.133 & 13.85 \\
SKSH1 & SLL3 & 12.494  & 7.28 & 10.08 & 12.09 & 1.541 & 8.26 \\
\hline
\end{tabular}
\end{center}
\vskip 0.2cm

\begin{center}
\begin{flushleft}
$\underline{\mbox{{\sf d) }\rm NN force=SV}}$ 
\smallskip \par [in $npe$ matter in $\beta$-equilibrium: 
$n_B (c_s^2=1)=4.983\ n_{\rm sat}$, $n_c (1.4\ M_\odot)=
2.10\ n_{\rm sat}$, $R (1.4\ M_\odot)$=13.46 km, $n_{\rm max}=
5.44\ n_{\rm sat}$, $M_{\rm max}=2.426\ M_\odot$, 
$R(M_{\rm max})=11.54\ $km]  
\end{flushleft}
\vskip 0.2cm
\begin{tabular}{|c|c||c||c|c||c|c|c|}
\hline
N$\Lambda$  & $\Lambda\Lambda$ & $n_{c_s^2=1}$ & $n_c (1.4\ M_\odot)$ 
& $R (1.4\ M_\odot)$ & $n_{\rm max}$ & $M_{\rm max}$ & $R(M_{\rm max})$ \\
\hline
LY-I  & no $\Lambda\Lambda$ & causal &   -  &   -   &  4.08 & 1.368 & 12.80 \\
LY-I  & SLL1  & $>20$  &   -  &   -   &  3.53  & 1.291 & 13.15 \\
LY-I  & SLL2  & 15.412 & 9.39 & 9.44  & 12.38 & 1.424 & 8.37 \\
LY-I  & SLL3  & 9.553  & 3.54 & 12.69 &  9.03   & 1.768 & 9.34 \\
\hline
LY-IV & no $\Lambda\Lambda$ & causal &   -  &   -  &  4.08 & 1.379 & 12.96 \\
LY-IV & SLL2  & 15.317 & 8.78 & 9.73 & 12.14 & 1.432 & 8.45  \\
\hline
YBZ1  & SLL2  & 13.291 & 2.52 & 13.35 & 9.57 & 1.544 & 9.52 \\
YBZ1  & SLL3  & 7.558  & 2.49 & 13.40 & 7.69 & 1.925 & 9.97 \\
\hline
YBZ6  & no $\Lambda\Lambda$ & causal & 2.19 & 13.40 & 4.44 & 1.662 & 12.75 \\
YBZ6  & SLL1  & $>20$  & 2.21 & 13.40 & 4.01 & 1.562 & 13.04 \\
YBZ6  & SLL2  & 13.082 & 2.22 & 13.41 & 6.5  & 1.639 & 11.24 \\
YBZ6  & SLL3  & 7.253  & 2.25 & 13.42 & 7.24 & 1.984 & 10.27 \\
\hline
SKSH1 & no $\Lambda\Lambda$ & $>20$  &   -  &   -   & 3.22  & 1.157 & 13.17 \\
SKSH1 & SLL2  & 18.318 &   -  &   -   & 3.06  & 1.111 & 13.24 \\
SKSH1 & SLL3  & 10.914 & 6.09 & 10.54 & 11.07 & 1.612 & 8.44  \\
\hline
\end{tabular}
\end{center}
\vskip 0.2cm
\begin{center}
{\bf Table V}: Conditions for causality and neutron star properties
\end{center}

\begin{figure}[htb]
\mbox{%
\parbox{10cm}{\epsfig{file=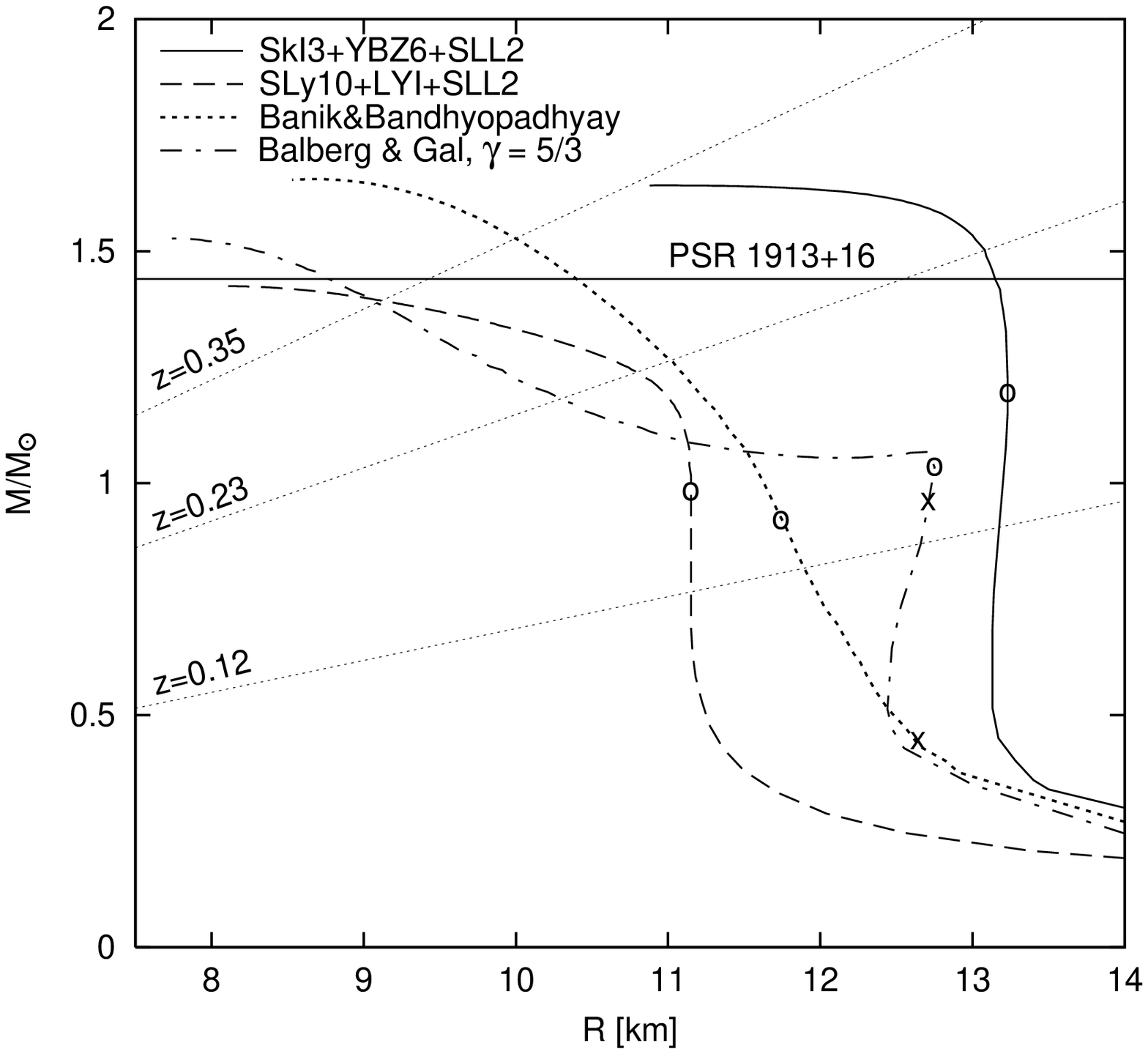,width=10cm}}
\parbox{7cm}{{\bf Fig. 6} -- Neutron star mass-radius relation for several
non relativistic models of baryonic matter.}
}
\end{figure} 

In Fig. 6 we compare the mass-radius relation obained in several
non relativistic models of matter in $\beta$-equilibrium with hyperons.
As discussed above, among our Skyrme parametrizations the best are the 
SkI3+YBZ6+SLL2 and the SLy10+LYI+SLL2 sets. The SkI5 or SV forces again
give results very similar to those of the SkI3 force. The maximum mass
reached with the stiffer SkI3+YBZ6+SLL2 set is 1.64 $M_\odot$ comfortably 
above the  Hulse-Taylor value 1.44 $M_\odot$ while the maximum mass 
permitted by the softer SLy10+LYI+SLL2 set, 1.42 $M_\odot$, falls a bit 
short. This is however not uncommon a feature in non relativistic models
(see Brueckner-Hartree-Fock calculations \cite{VPREHJ00,BBS00}).
Among the parametrizations given by Balberg and Gal \cite{BG97}, that 
with intermediate compressibility corresponding to $\gamma=5/3$ was chosen. 
Figure 6 also displays the results obtained with a density-dependent 
Seyler-Blanchard potential as parametrized by Banik and Bandhyopadyay 
\cite{BB00}. In contrast to our Skyrme forces the models of Balberg and Gal 
and of Banik and Bandhyopadhyay both take into account $\Sigma$ hyperons.
The threshold for hyperon formation is represented on the figure
with circles for the $\Lambda$ and crosses for the $\Sigma$.

\begin{center}
\begin{tabular}{|c|cc|cc|ccc|}
\hline
Model & $n_{\rm thr}^\Sigma$ & $n_{\rm thr}^\Lambda$ & $R(1.4\ M_\odot)$&
$n_c(1.4\ M_\odot)$ & $M_{\rm max}$ & $R(M_{\rm max})$ & $n_c(M_{\rm max})$ \\
\hline
Balberg $\gamma=5/3$ & 1.82 & 2.32 & 9.02 & 8.5 & 1.53 & 7.74 & 12.8 \\
Banik\& Bandhyo.     & 1.52 & 2.82 & 10.6 & 5.4 & 1.66 & 8.68 & 10.2 \\
\hline
\end{tabular}
\end{center}

The parametrization of Balberg and Gal is in fact unstable slightly above 
the $\Lambda$ hyperon threshold. As a consequence, the $M$ {\it vs.} $R$
relation displays an extended plateau, then recovers, crosses the line
$M=1.4\ M_\odot$ and finally reaches a maximum at $M=1.53\ M_\odot$.

Also represented on the figure are lines of constant gravitational redshift
\beq
z = \left(1 -{2 G M\over R c^2} \right)^{-1/2} -1 \nonumber
\eeq
To date two determinations exist from the observation of spectral lines
in isolated neutron stars \cite{redshift}. Sanwal {\it et al.} obtained
a result with a large error bar  $z=0.12 - 0.23$ while Cottam {\it et al.}
could extract a precise determination $z=0.35$.  All the neutron star 
models displayed would be compatible with the determination of Sanwal 
{\it et al.}. Almost all models are also in agreement with the value
$z=0.35$ of Cottam {\it et al.}. On the other hand, the mass-radius 
relation from the SkI3+YBZ6+SLL2 set is only marginally compatible 
with this value.

Figure 7 displays the density profile of a 1.4 $M_\odot$ neutron star 
and its hyperonic content. The profiles corresponding
to two of the Skyrme parameters sets studied in this work, SkI3+YBZ6+SLL2 
and SLy10+LYI+SLL2 (top panels) are compared to a version of the models 
of Balberg and Gal and of Banik and Bandhyopadhyay (medium and bottom 
left panels) where all hyperons save the $\Lambda$s are artificially 
switched off. The profiles corresponding to the latter two models when 
also $\Sigma^-$ hyperons are taken into account is displayed on the medium 
and bottom right panels. We can see that additional hyperons make the 
equation of state softer and the neutron star more compact.

\begin{figure}[htb]
\mbox{%
\parbox{7.3cm}{\epsfig{file=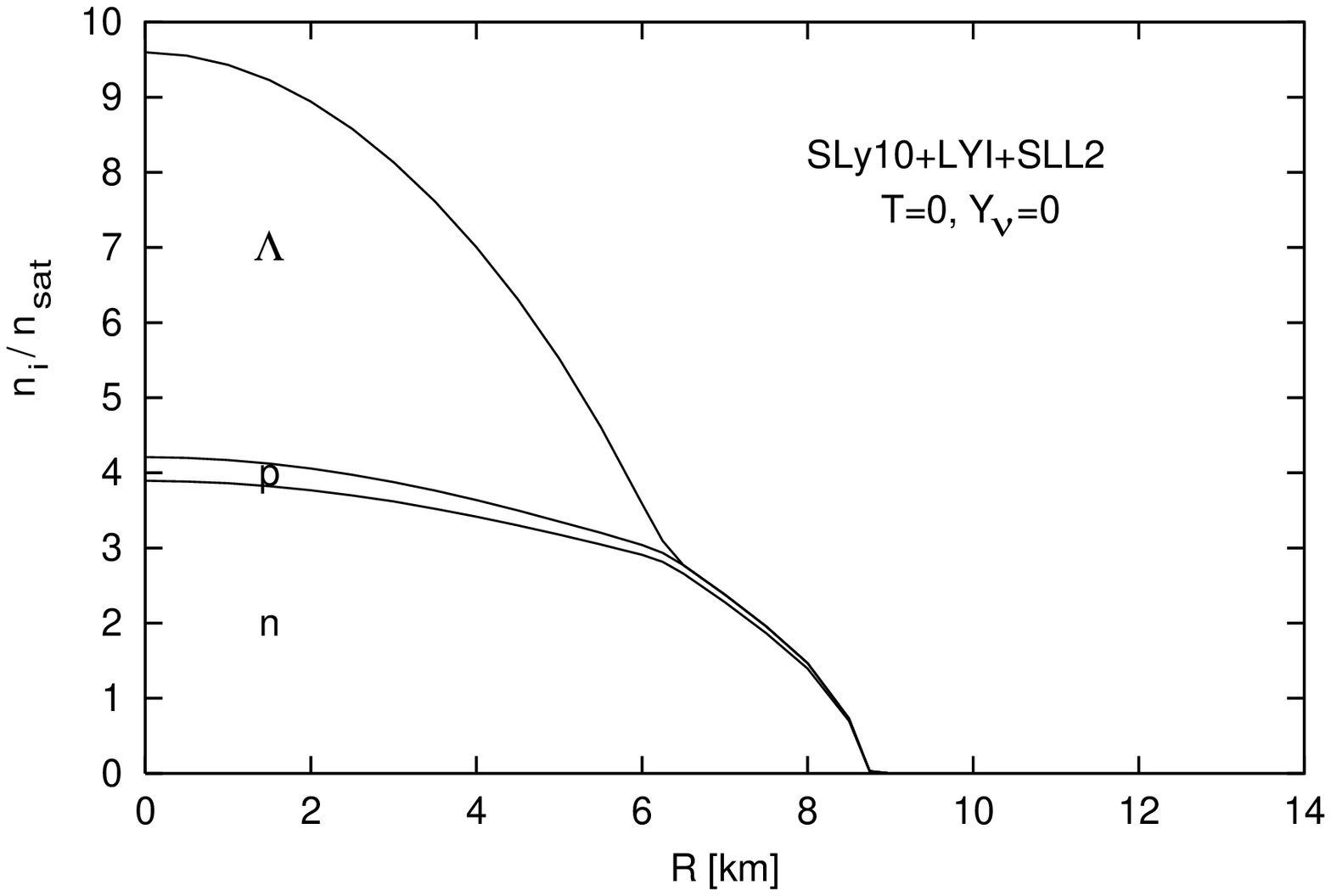,width=7.1cm}}
\parbox{0.5cm}{\phantom{aaaa}}
\parbox{7.3cm}{\epsfig{file=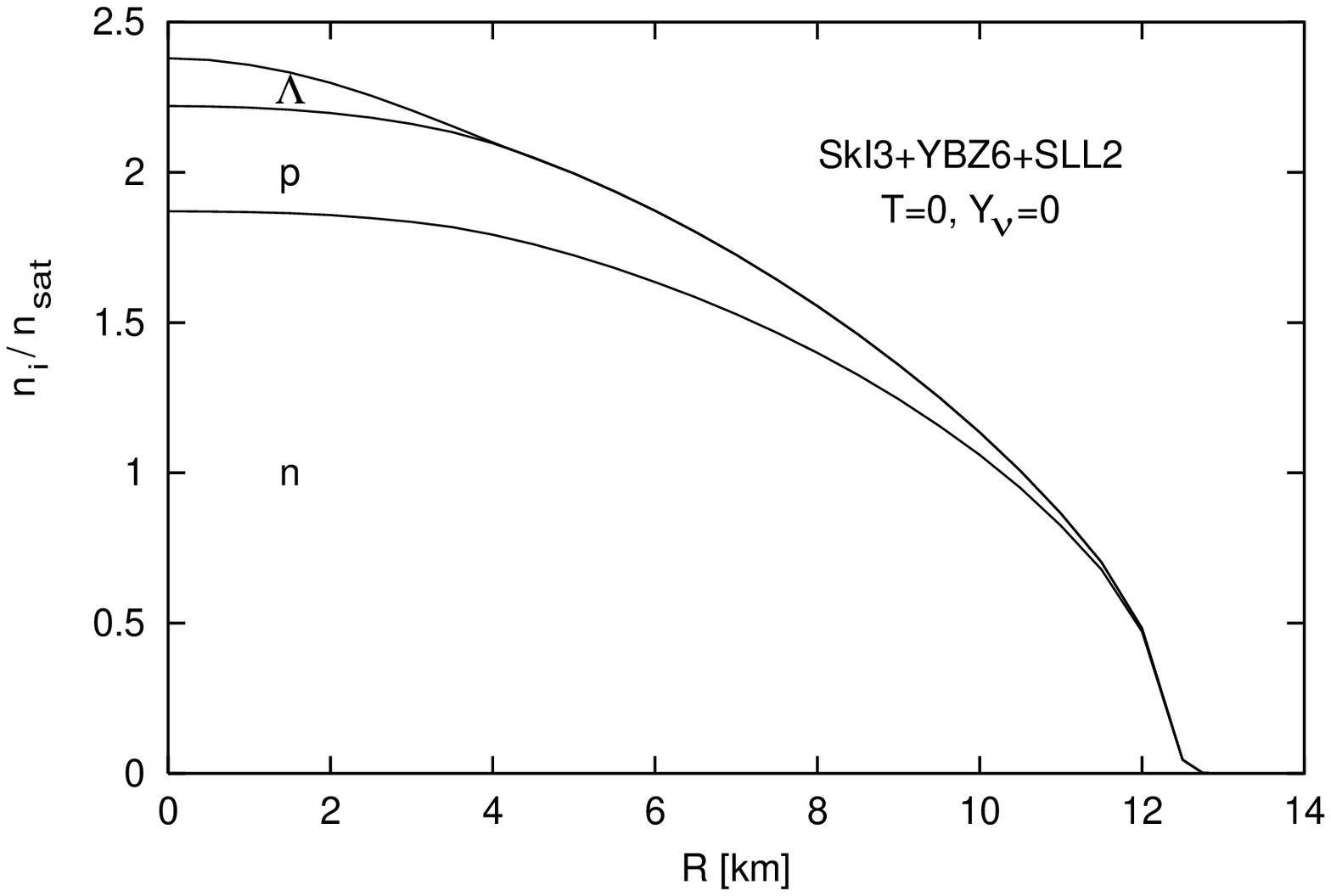,width=7.1cm}}
}
\mbox{%
\parbox{7.3cm}{\epsfig{file=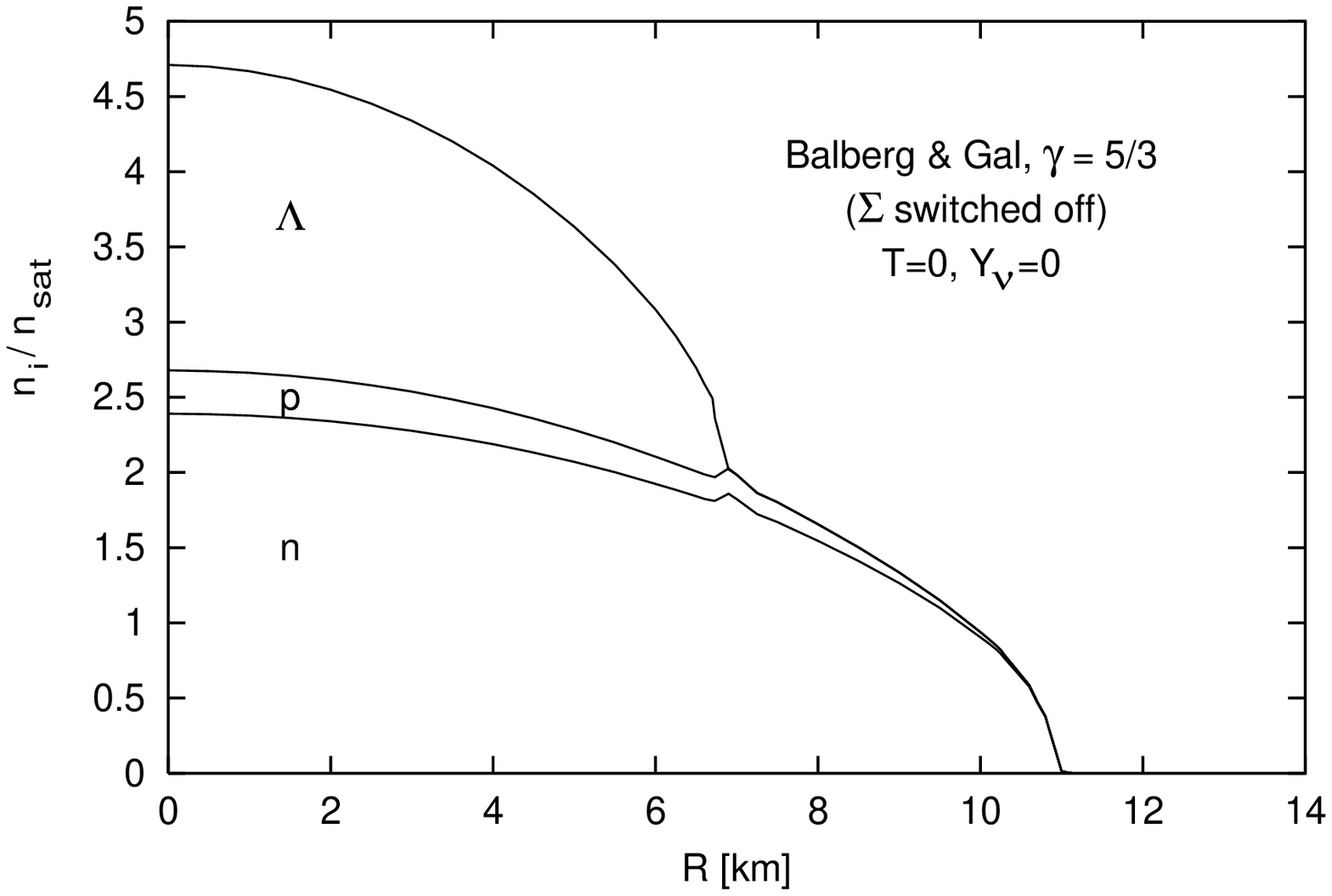,width=7.1cm}}
\parbox{0.5cm}{\phantom{aaaa}}
\parbox{7.3cm}{\epsfig{file=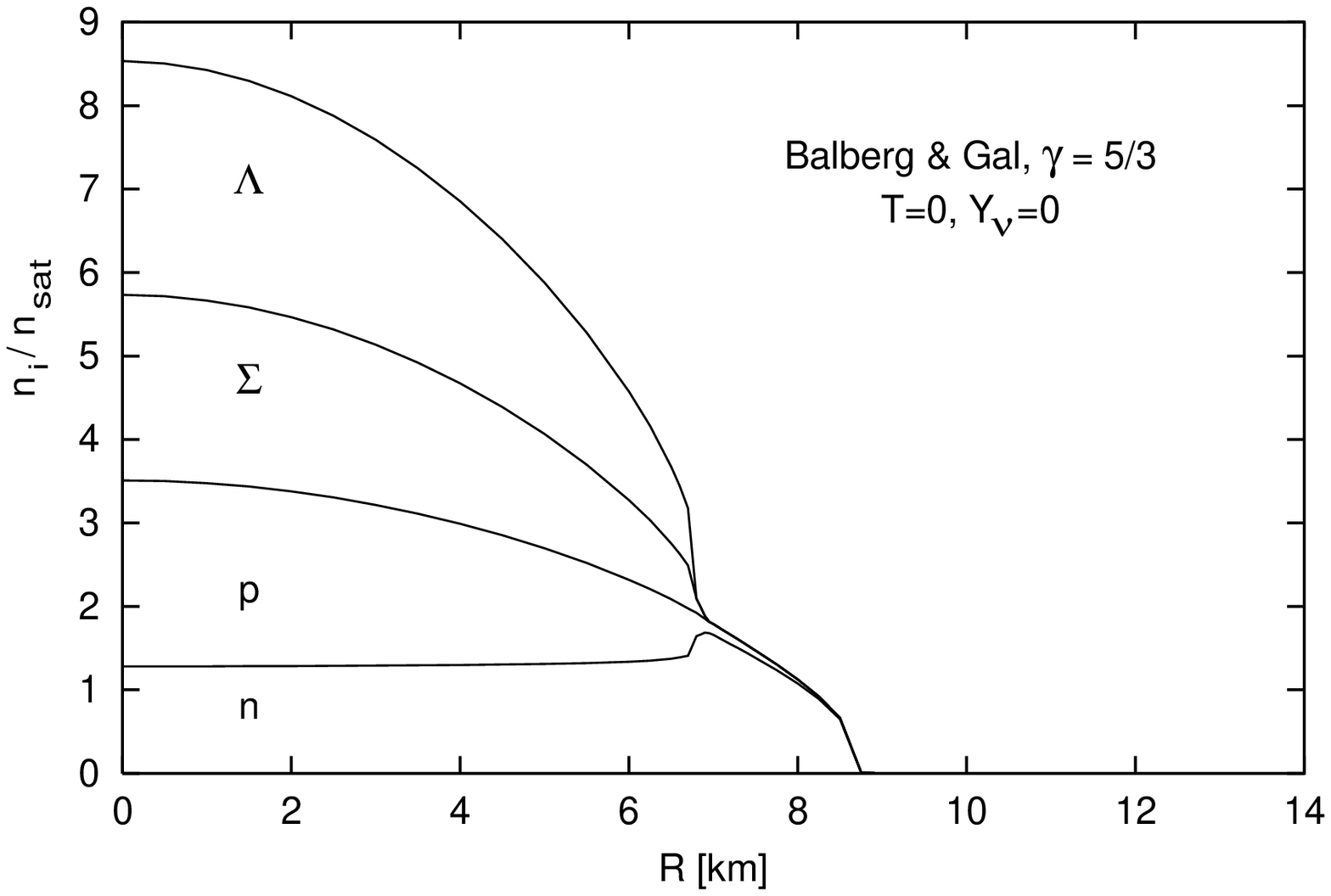,width=7.1cm}}
}
\mbox{%
\parbox{7.3cm}{\epsfig{file=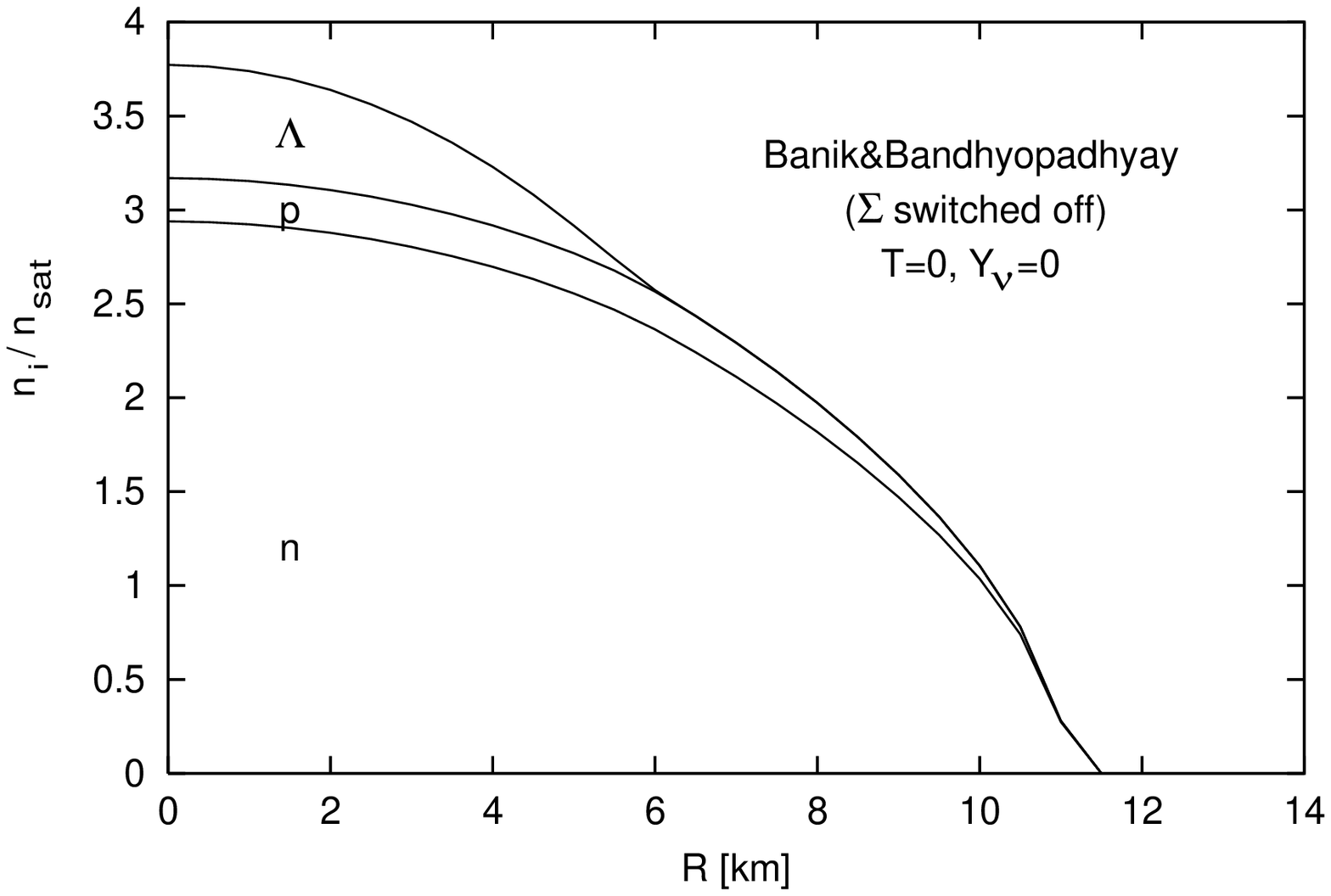,width=7.1cm}}
\parbox{0.5cm}{\phantom{aaaa}}
\parbox{7.3cm}{\epsfig{file=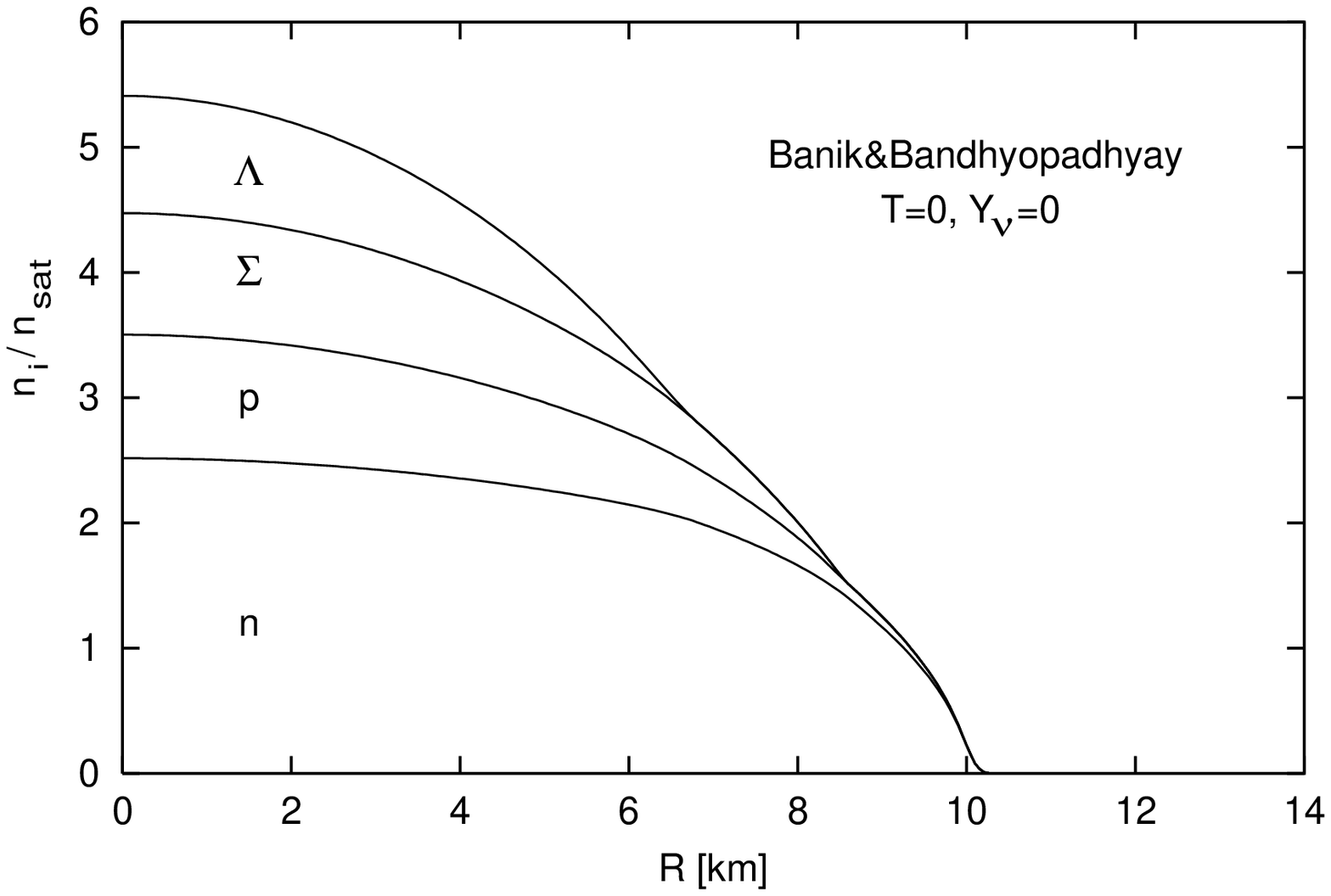,width=7.1cm}}
}
\vskip 0.3cm
\mbox{%
\parbox{16cm}{{\bf Fig. 7} -- Density profile of a 1.4 $M_\odot$ neutron star
for several non relativistic models of baryonic matter.}
}
\end{figure}

\section{Finite temperature effects, neutrino trapping}
\label{S:finiteT}

According to proto-neutron star formation and cooling calculations 
(see {\it e.g.} \cite{PRPLM99}), temperatures of the order of 30 to 50 
MeV can be reached in the late phases of the supernova collapse. 
We therefore investigate here the range $T \in [0-50]$ MeV. 

The thermodynamical quantities are now to be written in terms of Fermi 
integrals. The expressions which are given in the Appendix at $T=0$ are 
easily generalized: The densities $\rho_i$ and the quantities $\tau_i$ 
related to the kinetic energy should be replaced in the expressions for 
the baryonic contributions to the energy density and the effective masses 
by (in units $\hbar=c=k_B=1$):
\beq
\rho_i &=& {1 \over 2 \pi^2} \left( 2 m_i^* T \right)^{3/2} I_{1/2}(\eta_i) 
   \quad , \qquad \tau_i = {1 \over 2 \pi^2} \left( 2 m_i^* T \right)^{5/2}
   I_{3/2}(\eta_i) \nonumber \\
   {\rm with} && I_n= \int_0^\infty {u^n du \over 1+ e^{u-\eta_i}}
\nonumber
\eeq
while the chemical potentials is related to the $\eta_i$ by $\mu_i = \eta_i T 
+ {\cal U}_i(\rho_i,\tau_i)$. The pressure is obtained from the derivative 
of the free energy 
\beq
{\cal F} = {\cal E} -T {\cal S} \quad , \qquad 
{\cal E}=\sum_{A,B=N,\Lambda}{\cal E}_{AB} + {\cal E}_{\rm leptons}
\quad , \qquad 
P = \rho^2 {\partial^2 ({\cal F}/\rho) \over \partial \rho^2} 
\nonumber
\eeq
with the entropy 
\beq
S=\sum_{i=n,p,\Lambda} S_i +S_{\rm leptons}
\ ,\quad S_i = {5 \tau_i \over 6 m_i^* T} -\rho_i\, \eta_i
\nonumber
\eeq

We used the GFD\_D3 code published by Gong {\it et al.} \cite{CPC136} 
to calculate the Fermi integrals and keep as before the leptons fully 
relativistic whereas the baryons are treated nonrelativistically  
in consistency with the use of the Skyrme interaction.

\begin{figure}[htb]
\mbox{%
\parbox{8cm}{\epsfig{file=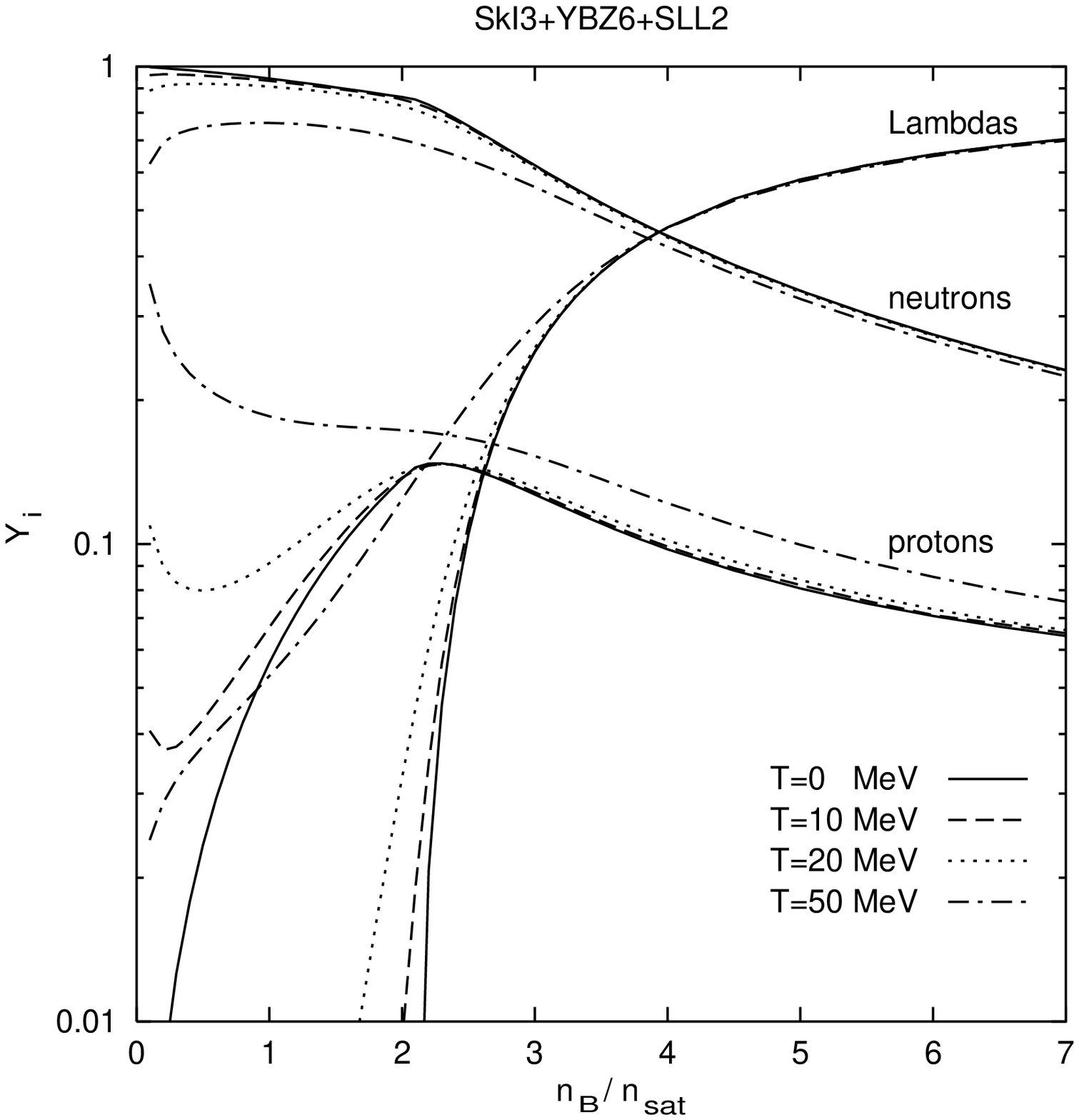,width=8cm}}
\parbox{8cm}{\epsfig{file=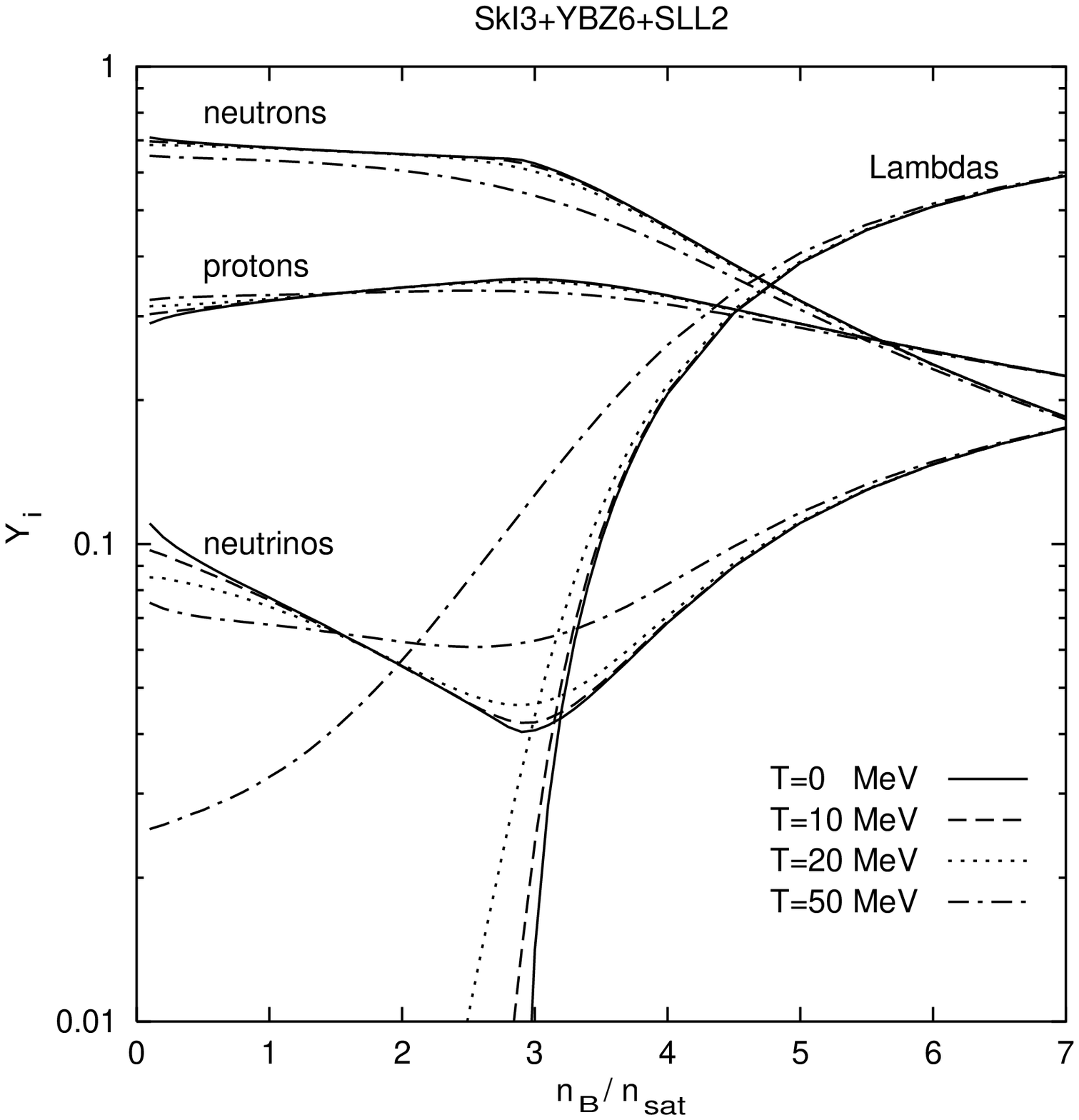,width=8cm}}
}
\vskip 0.2cm
\mbox{%
\parbox{16cm}{{\bf Fig. 8} -- Composition of neutron star matter in
$\beta$ equilibrium at finite temperature, calculated with the 
parametrization SkI3+YBZ6+SLL2 - (a) without neutrino trapping - 
(b) with neutrino trapping, $Y_L=0.4$}
}
\end{figure}

The effect of temperature on the effective masses begins to be significant 
at $T \ge 30$ MeV; it tends to increase the neutron mass and decrease the
proton effective mass so that their difference is reduced. The presence of
trapped neutrinos, which tends to render the matter less asymmetric, 
also reduce the proton-neutron mass difference. In a warm protoneutron 
star, both effects are present and cumulate, the main contribution coming 
from $Y_\nu \ne 0$. The hyperon mass on the other hand is not significantly
modified by temperature or neutrino trapping.

The chemical composition is plotted on Fig. 8, without neutrino trapping
on the left panel and with neutrino trapping on the right panel, for
four values of the temperature, T=0, 10, 20 and 50 MeV. We used in
this section the parametrization SkI3 + YBZ6 + SLL2. 

Let us first discuss the neutrino-free case. It can be seen on Fig. 8-a 
that finite temperature effects are more important at moderate densities 
$n_B < 3 n_{\rm sat}$. Until $T=1$ MeV the results are undistinguishable 
from the T=0 case. When T is increased from 1 to 50 MeV, the matter is 
more symmetric ({\it i.e} the proton fraction increases). There is
{\it stricto sensu} no threshold for $\Lambda$ hyperon production anymore, 
rather there always exist a vanishingly small number of $\Lambda$s for 
any value of the baryonic density below the $T=0$ threshold. Nevertheless, 
at the temperatures considered here, it is still possible to define a 
threshold density for practical purposes, which then moves to lower values 
as the temperature increases.

The equation of state is somewhat stiffer at finite $T$, but this
hardly affects the structure of the star until $T$=20 MeV. This result
concern the case when no neutrinos are present in the matter. It is known
however that high temperature shorten drastically the mean free path 
of the neutrinos so that a large amount of $\nu$ are trapped. A typical 
value resulting from supernova collapse calculations is that the lepton 
fraction is of the order of $Y_L\simeq 0.4$. Fig. 8-b was drawn assuming 
this value for $Y_L$.

As expected the matter is more symmetric when neutrinos are trapped
and the threshold for hyperon production is shifted to higher density.
For example with the SkI3+YBZ6+SLL2 parametrization and at $T=0$
we have ($n_{\rm thr}=2.08$, $Y_p^{\rm thr}=0.14$) for $Y_\nu=0$ 
and ($n_{\rm thr}=2.84$, $Y_p^{\rm thr}=0.36$) for $Y_L=0.4$.
For a given lepton fraction $Y_L$, varying the temperature has even
less impact on the composition of neutrino-trapped matter than in
neutrino-free matter. Fig. 9-a shows that the neutrino trapped matter 
with hyperons is stiffer than its neutrino-free counterpart, a known 
result which leads to interesting consequences regarding a possible 
category of metastable stars which would collapse to a black hole as 
they cool when the deleptonization phase is completed (see {\it e.g.}
\cite{PBPELK97}). Fig. 9-b  illustrates this feature for the parameter 
set SkI3+YBZ6+SLL2: A newly formed protoneutron star with $T=30$ MeV 
and $Y_L=0.4$ is metastable if its mass lies in the range $M \in 
[1.64-1.88] \ M_\odot$ (and radius in the range $R \in [12.08-14.07]\  
{\rm km}$). For a lower mass, the star contracts as it looses its 
neutrinos and cools and its hyperonic content increases.

\begin{figure}[htb]
\mbox{%
\parbox{8cm}{\epsfig{file=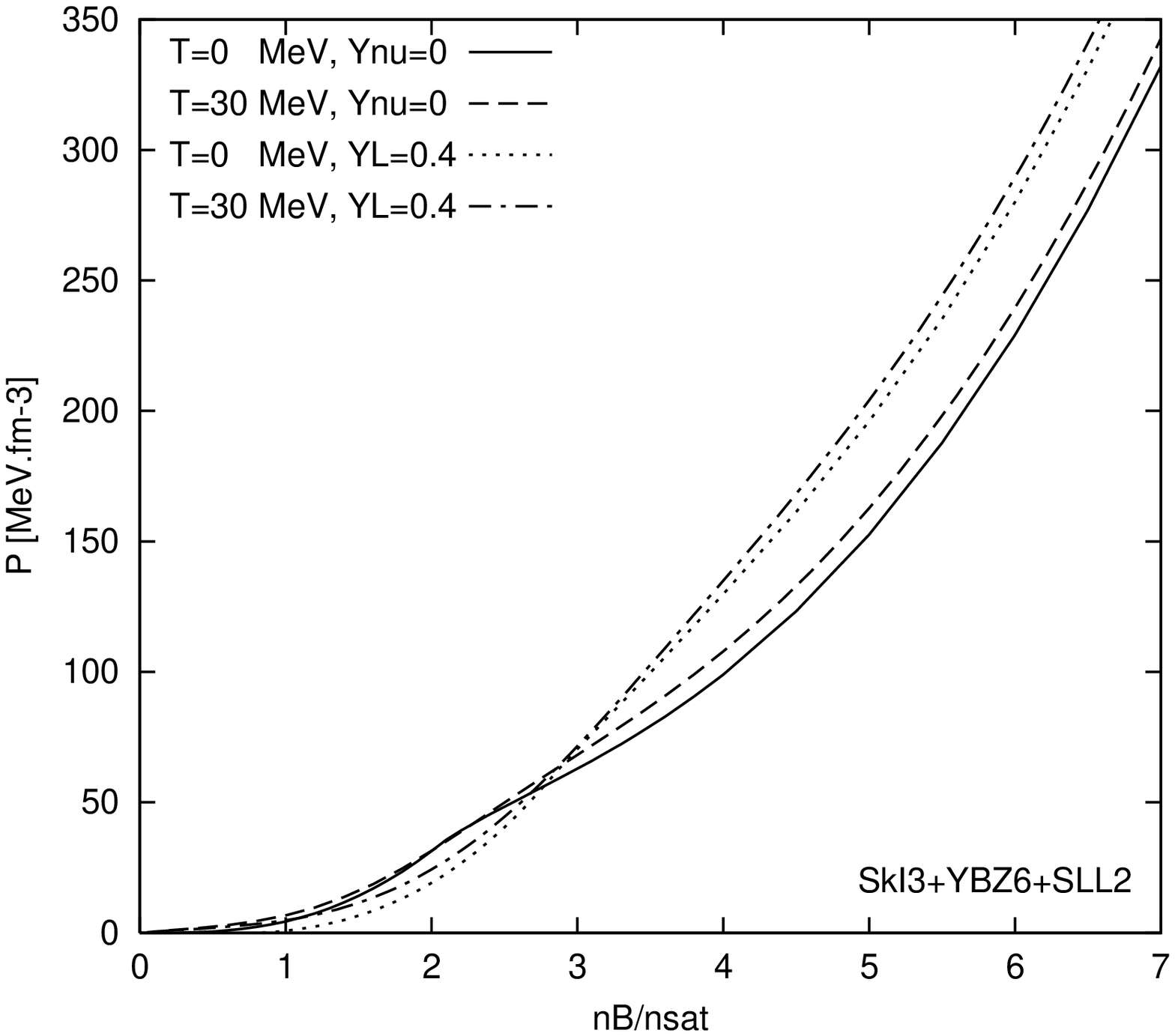,width=8cm}}
\parbox{8cm}{\epsfig{file=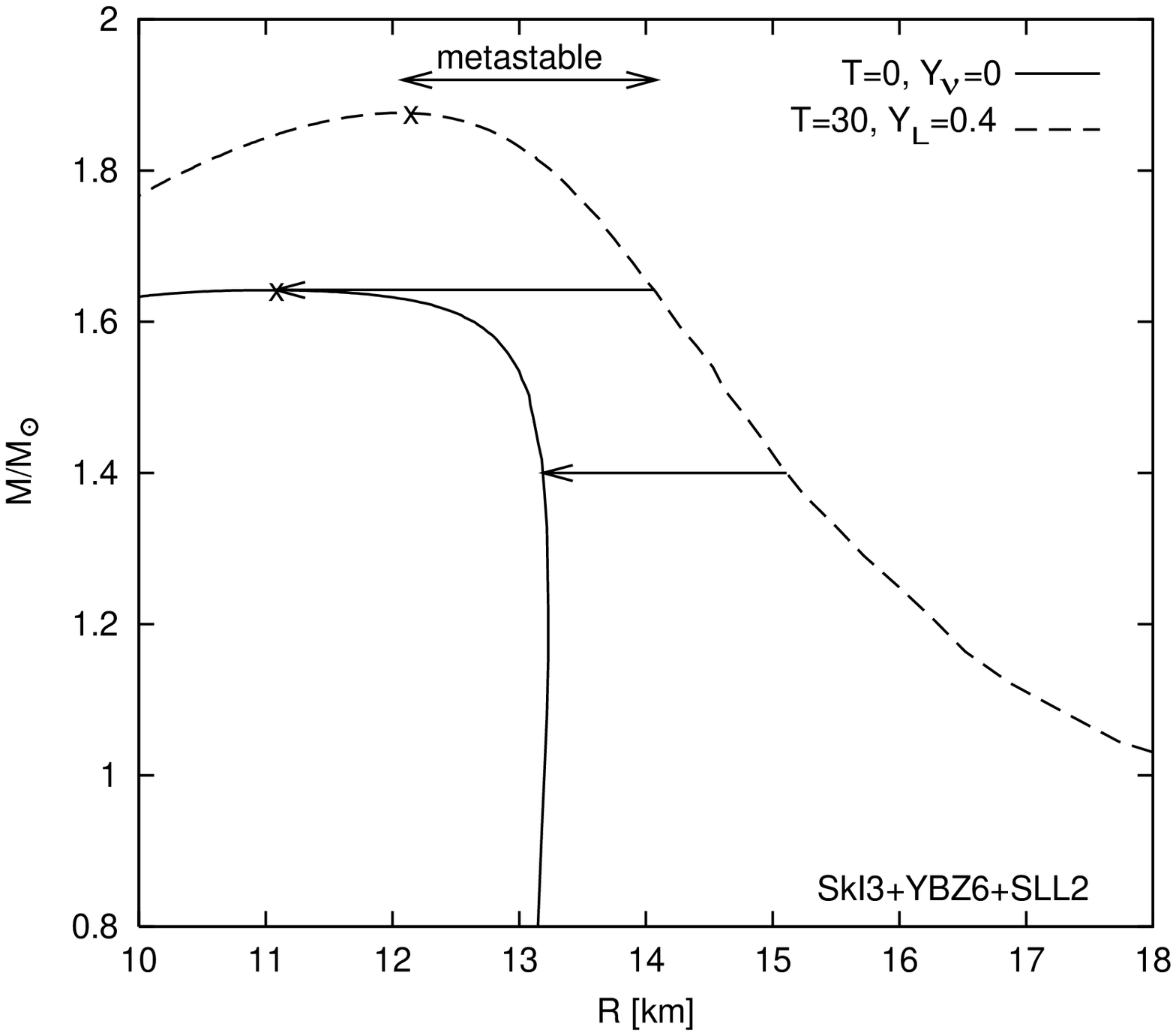,width=8cm}}
}
\vskip 0.2cm
\mbox{%
\parbox{16cm}{{\bf Fig. 9} --  Effect of neutrino trapping and finite
temperature on (a) the equation of state (b) the structure of the 
neutron star}}
\end{figure}

\begin{center}
\begin{tabular}{|l||c|c||c|c||c|c|c|}
\hline
SkI3+YBZ6+SLL2 & $n_{\rm ferro}$ & $n_{c_s^2=1}$ & $n_c (1.4\ M_\odot)$ 
& $R (1.4\ M_\odot)$ & $n_{\rm max}$ & $M_{\rm max}$ & $R(M_{\rm max})$ \\
\hline
$T=0$, $Y_\nu=0$      & 9.19  & 13.08 & 2.39 & 13.19 & 6.79 & 1.64 & 11.02\\
$T=30$ MeV, $Y_L=0.4$ & 11.37 & 12.02 & 2.66 & 15.11 & 6.38 & 1.88 & 12.08\\
\hline
\end{tabular}
\end{center}

As a conclusion to this subsection, our results concerning the effects of
temperature and neutrino trapping are in full agreement with those obtained 
with other models of the baryonic interaction. 
The influence of the temperature on the equation of state and chemical 
equilibrium comes mostly indirectly through the buildup of an important 
fraction of trapped neutrinos.

\section{Conclusion}
\label{S:ccl}

The aim of this work was threefold:
\par\indent 
{\bf -} Use the most recent Skyrme parametrizations with hyperons 
existing on the market which are adjusted to reproduce the data on 
nuclei and hypernuclei and test them on neutron stars, 
\par\indent 
{\bf -} Study the influence of the hyperons on the ferromagnetic 
transition,
\par\indent 
{\bf -} Inquire whether the hyperons are likely to affect the tail of 
the neutrino burst in supernova explosions and prepare the background 
for the calculation of the neutrino mean free path in protoneutron stars.

\medskip

While the model considered in this work is still very schematic, 
it has led to several interesting results. 

The first result is that the presence of hyperons generally delay the 
onset of the ferromagnetic instability, and in many cases they are even
able to remove it completely.

An other advantage is that, by softening the equation of state, 
the hyperons remove the causality flaw and keep $c_s^2 <1$ even
up to very large densities.

It is rather encouraging that NN and N$\Lambda$ interactions from 
different authors and apparently very dissimilar parameter sets 
(compare {\it e.g.} SV to SkI3 or SKSH1 to LY-I in Tables I, II !), 
once a series of reasonable requirements are fulfilled, yield very 
similar results qualitatively and even quantitatively. The major 
incognita is the $\Lambda$-$\Lambda$ interaction which was very 
poorly known at the time the interactions used in this work were 
designed.  

After studying 43 NN forces in combination with 13 parametrizations 
of the N$\Lambda$ force and 4 options for the $\Lambda\Lambda$ forces,
all taken from the litterature and known to reproduce correctly the 
properties of nuclei and hypernuclei, four combinations of  
NN+N$\Lambda$+$\Lambda\Lambda$ parameter sets were selected:
(SkI3+YBZ6+no $\Lambda\Lambda$), (SkI3+YBZ6+SLL2), (SV+YBZ6+SLL2) 
and (SLy10+LYI+SLL2). Replacing SkI3 by SkI5 or LYI by LYIV in the 
above sets would only give slightly different results. These sets 
fulfill all of the following conditions: 
\par\indent {\it (i)} The NN force belongs to the subset selected by 
Rikovska Stone {\it et al.}. The effective masses behave smoothly in 
all the relevant density and temperature range, pure neutron matter 
is always stable, the asymmetry energy does not decrease so much with 
density  that protons would disappear from the system, neutron stars
formed from $npe$ matter in $\beta$ equilibrium can reach a mass at 
least equal to 1.4 $M_\odot$
\par\indent {\it (ii)} The N$\Lambda$ force belongs to the set preferred 
by Lanskoy {\it et al.} as best reproducing the properties of hypernuclei.
The threshold for hyperon formation should lay between 1.7 and 4 times 
saturation
\par\indent {\it (iii)} The neutron star formed of softer $np\Lambda e$ 
matter in $\beta$ equilibrium should still reach a mass at least equal 
to 1.4 $M_\odot$. 
\par\indent {\it (iv)} No ferromagnetic transition should be present
As explained in the main text, this is, for a Skyrme parametrization, 
a strong requirement.

The second result of this work is that the selected sets are also found 
to reproduce all features observed in other models such as non-relativistic 
Brueckner-Hartree-Fock or relativistic mean field calculations, not only
qualitatively but also quantitatively. The threshold for hyperon formation 
is in fact restricted to the narrower range [2--2.5] $n_{\rm sat}$. The 
softening of the equation of state brings the maximum mass of the star 
from 2 -- 2.5 $M_\odot$ for a npe star down to 1.4 -- 1.6 $M_\odot$ for 
a npYe star. This has even lead to some speculation (see {\it e.g.} 
\cite{VPREHJ00}) whether the clustering of known pulsar masses around the 
value 1.4 $M_\odot$ would be due to this feature of the equation of state 
rather than from the circumstances of their formation in a supernova.
The metastability of hot stars with trapped neutrinos when hyperons
are present, as discussed {\it e.g.} in \cite{PBPELK97}, is also recovered.

\medskip

The selected sets are also applied in a companion paper \cite{M04b} 
to the calculation of the scattering rate of neutrinos in baryonic matter,
for a finite temperature and non vanishing number of trapped neutrinos.
The possibility to access to the properties of polarized matter was a 
further motivation to this work: It was necessary to be able to calculate 
the Landau parameters in the spin $S=1$ channel, since the axial channel 
is dominant for neutrino scattering.

As a protoneutron star is formed in a supernova explosion, the high amount
of neutrinos trapped in its interior by high temperature and density gradually
diffuse out in the first 50 seconds as the star cools. Hyperons begin to 
appear in the course of this deleptonization process and could affect the 
tail of the neutrino signal. Modern detectors would now be able to detect
this effect. The idea was explored by Reddy {\it et al} \cite{RPL98} in 
the mean field approximation. The calculation is performed in the random phase
approximation in \cite{M04b}.  The parameter sets selected in the present 
work predict a non negligible hyperon content for $t \ge 20\ s$ after
the collapse (see Figs 2 and 3 of \cite{M04b}).

\medskip

Let us examine again the main shortcomings of the present model and the 
way it could be improved in future work. 

\par\indent {\it (i)} 
The Skyrme interaction is long known for displaying a series of problems 
at high density: behavior of asymmetry energy, onset of ferromagnetism, 
causality ... While we were able to avoid these problems by carefully 
passing the existing parametrizations through the crib of these various 
constraints, we cannot avoid the feeling that we are pushing the Skyrme 
model much beyond its capacity. More refined models do not have these
problems; for example the non relativistic Brueckner Hartree Fock 
calculations do not show a ferromagnetic transition. 
A possible answer to this point would be to develop a parametrization
of Brueckner-Hartree-Fock calculations in terms of an energy density 
functional for polarized matter with a non vanishing hyperonic content.
Some steps have already been performed in this direction by Vida\~na 
{\it et al.}. These authors studied the polarized neutron-proton matter 
system \cite{VBPR02} and concluded to the absence of a ferromagnetic 
transition. They also obtained the equation of state of unpolarized
npY matter in $\beta$ equilibrium and applied it to the calculation of 
neutron star mass radius relation \cite{VPREHJ00} and parametrized the 
neutron-proton-Lambda system \cite{VPRS01} for applications to hypernuclei.
The parametrization of Brueckner-Hartree-Fock results with modern Nijmegen 
potentials for the full polarized $np\Lambda\Sigma$ are currently 
under way \cite{ournext}.

\par\indent {\it (ii)} We have seen that the effect of the 
$\Lambda$-$\Lambda$ force at high density is very important,
whereas the description of the hyperon-hyperon interaction in
phenomenological formalisms is still very poor. Brueckner-Hartree-Fock 
calculations offer a coherent framework to calculate the hyperon-hyperon
interaction in medium starting from bare potentials such as the Nijmegen
one known from scattering data of free particles, and then cross checking 
the results with the data on double hypernuclei. This has actually been
performed with the Nijmegen NSC97e model {\it e.g.} in \cite{VRP03}. 

\par\indent {\it (iii)} Relativistic effects can be expected to play
an important role at high density. Many of the problems encountered
in the Skyrme parametrization mentioned above in this paragraph
(behavior of asymmetry energy, onset of ferromagnetism and causality)
are not present in the relativistic formulation. Relativistic mean field 
models with hyperons (see {\it e.g.} \cite{RMF-hyps}) are so well 
under control that they have made their way into textbooks \cite{Gl-book}. 
The parametrizations of the present work obviously do not pretend to 
compete with this line of work at the mean field level; rather they were 
developped in view of their application at the RPA level where they give 
rise to simpler Dyson equations than in the relativistic formulation. 
The relativistic extension at RPA level to hyperons would be completely 
straightforward but somewhat unwieldy.

While it could be argued that the physical foundations of the Skyrme 
model are disputable when applied in the context of neutron stars, 
the parametrizations presented in this paper should rather be considered 
as reliable phenomenological models for run-of-the-mill calculations.
Once the order of magnitude of physical effects are ascertained with this
simple model, they can refined by applying more complicated 
Brueckner-Hartree-Fock or/and relativistic models.

\vskip 0.5cm
\noindent {\large{\bf Acknowledgements}}
\vskip 0.2cm

This work was supported  by the spanish-european (FICYT/FEDER) 
grant number  PB02-076. Part of it was realized during a stay at the 
Departament d'Estructura i Constituents de la Materia of Barcelona 
University. Several discussions with A. Polls are gratefully acknowlegded.

\newpage

\section{Appendix}

Besides the the usual parameterization of the nucleon-nucleon interaction
\beq
V_{NN}(r_1 -r_2) &=&  t_0\, (1+x_0 P_\sigma) \delta(r_1-r_2) 
+ {1 \over 2} t_1\, (1+x_1 P_\sigma) \left[ k'{}^2\  \delta(r_1-r_2) 
+ \delta(r_1-r_2)\ k^2 \right] \nonumber \\
& & \mbox{\hskip -0.8cm}
+ t_2\, (1+x_2 P_\sigma)  k'\, \delta(r_1-r_2)\, k + {1 \over 6} t_3\, 
(1+x_3 P_\sigma)\, \rho_N^\alpha \left({r_1+r_2 \over 2}\right) 
\delta(r_1- r_2) , 
\eeq
we use the following Lambda-nucleon and Lambda-Lambda potentials
\cite{LY97,L98} :
\beq
V_{N\Lambda} (r_N-r_\Lambda) &=& 
u_0\, (1+y_0 P_\sigma) \delta(r_N-r_\Lambda) + {1 \over 2} u_1\, 
\left[ k'{}^2\  \delta(r_N-r_\Lambda) +  \delta(r_N-r_\Lambda)\ k^2 \right] 
\nonumber \\
&& \mbox{\hskip -0.8cm}
+ u_2\, k'\, \delta(r_N-r_\Lambda)\, k + {3 \over 8} u_3\, (1+y_3 P_\sigma) 
\rho_N^\beta \left({r_N+r_\Lambda \over 2}\right)  \delta(r_N-r_\Lambda) \\
V_{\Lambda\Lambda} (r_1-r_2) &=& \lambda_0 \delta(r_1-r_2) +{1 \over 2}
\lambda_1 \left[ k'{}^2\ \delta(r_1-r_2) + \delta(r_1-r_2)\ k^2 \right] 
\nonumber \\
&& \mbox{\hskip -0.8cm}
+ \lambda_2\, k'\, \delta(r_1- r_2)\, k + \lambda_3\, \rho_\Lambda 
\rho_N^\gamma
\eeq
(We haved dropped in these expressions the spin-orbit terms which are
not used in this paper).
We obtain the energy density in  homogeneous matter in the usual way. 
In spin saturated $np\Lambda$ matter, the functional reads:
\beq
{\cal E} &=& <\psi\, |\, H \,|\, \psi > ={\cal E}_{NN} + {\cal E}_{N\Lambda} 
+ {\cal E}_{\Lambda\Lambda} \nonumber \\
{\cal E}_{NN} &=& {\hbar^2 \over 2 m_N} \tau_N + {t_0 \over 2}
\left[\left( 1+{x_0\over 2} \right)  \rho_N^2 - \left( {1 \over 2} + x_0 \right) 
(\rho_n^2 + \rho_p^2) \right] \nonumber \\
&& +{t_3 \over 12} \rho_N^\alpha \left[ 
\left( 1+{x_3\over 2} \right) \rho_N^2  -  \left( {1 \over 2} + x_3 \right) 
(\rho_n^2 + \rho_p^2) \right]  \\
&& + {1 \over 8} \left[ t_1 (2 + x_1) + t_2 ( 2+x_2) 
\right] \rho_N \tau_N -   {1 \over 8} \left[ t_1 (2 x_1 +1) - t_2 ( 2 x_2 +1)
 \right] (\rho_n\tau_n + \rho_p \tau_p ) \nonumber \\
{\cal E}_{N\Lambda} &=& u_0 \left( 1+{y_0\over 2} \right) \rho_N 
\rho_\Lambda +{3\over 8} u_3\,  \rho_N^{\beta+1} \rho_\Lambda \left( 1+{y_3\over 2} 
\right) \nonumber \\
&& + {1 \over 8}  \left[ u_1 (2 + y_1) + u_2 (2+y_2) \right]  
\left( \rho_N \tau_\Lambda + \rho_\Lambda \tau_N \right) \\
{\cal E}_{\Lambda\Lambda} &=&  {\hbar^2 \over 2 m_\Lambda } \tau_\Lambda 
+{\lambda_0 \over 4}  \rho_\Lambda^2 + {1 \over 8} (\lambda_1 + 3 \lambda_2)\,
\rho_\Lambda \tau_\Lambda +{\lambda_3 \over 4}\, \rho_\Lambda^2 \rho_N^\gamma
\eeq 
with $\rho_N=\rho_n+\rho_p$, $\tau_N=\tau_n+\tau_p$. At $T=0$ we have
$\tau_i=(3/ 5)\, \rho_i k_{Fi}^2$ and $\rho_i=k_{Fi}^3/(3 \pi^2)$.
At $T \ne 0$ these expressions should be replaced by Fermi integrals, 
see \S \ref{S:finiteT}. 
At $T=0$ we obtain the chemical potentials
\beq
\mu_n &=& {\partial{\cal E} \over \partial \rho_n} = 
{\hbar^2 \over 2 m_n^*} k_{Fn}^2 + {\cal U}_n(\rho_i,\tau_i) \nonumber \\
{\cal U}_n(\rho_i,\tau_i) &=& { t_0 \over 2} \left[
   \rho_n (1 -x_0) + \rho_p ( 2+x_0) \right] \nonumber \\
&& \mbox{\hskip -1.5cm}
   + {t_3 \over 24} \rho_N^{\alpha-1}\, \left[ \rho_n^2 \left\{ 
   (2+\alpha)(1-x_3) \right\} +  \rho_p^2 \left\{ 2\, (2+x_3)
   + \alpha (1-x_3)  \right\} + 2 \rho_n \rho_p \left\{3 + \alpha(2+x_3) 
   \right\} \right] \nonumber \\
&& \mbox{\hskip -1.5cm}
  +{1 \over 8} \left[ t_1(1-x_1) +3 t_2 (1+x_2) \right] 
   \tau_n + {1 \over 8} \left[ t_1(2+x_1) +t_2 (2+x_2) \right] \tau_p 
   + {\lambda_3 \over 4} \gamma \rho_\Lambda^2 \rho_N^{\gamma -1}  
   \nonumber \\
&& \mbox{\hskip -1.5cm}
   + u_0 \left( 1+{y_0 \over 2} \right) \rho_\Lambda 
   +{3 \over 8} u_3 (\beta+1) \left( 1+{y_3 \over 2} \right) \rho_N^\beta 
   \rho_\Lambda  + {1 \over 8} \left[ u_1 (2 + y_1) + u_2 (2+y_2) \right]  
   \tau_\Lambda 
\eeq
(The chemical potential for the proton $\mu_p$ can be obtained from $\mu_n$ 
by interchanging the indices $n$ and $p$) and 
\beq
\mu_\Lambda &=&  {\partial{\cal E} \over \partial \rho_\Lambda }=
  {\hbar^2 \over 2 m_\Lambda^*} k_{F\Lambda}^2 + 
   + {\cal U}_\Lambda(\rho_i,\tau_i) \nonumber \\
{\cal U}_\Lambda(\rho_i,\tau_i) &=&
   {\lambda_0 \over 2} \rho_\Lambda + {\lambda_3 \over 2} \rho_N^\gamma  
   \rho_\Lambda +{1 \over 8} (\lambda_1 + 3 \lambda_2) \tau_\Lambda 
   \nonumber \\
&& \mbox{\hskip -1.5cm}
+u_0 \left( 1+{y_0 \over 2} \right) \rho_N + 
   {3 \over 8} u_3  \left( 1+{y_3 \over 2} \right) \rho_N^{\beta+1} 
   +{1 \over 8} [ u_1 (2 + y_1) + u_2 (2+y_2) ]  
   \tau_N  
\eeq
The effective masses are given by
\beq
{\hbar^2 \over 2 m_n^*} &=&  {\hbar^2 \over 2 m_N} + {1 \over 8} 
   \left[ t_1 (2 + x_1) + t_2 (2+x_2) \right] \rho_N  -
   {1 \over 8} \left[ t_1 (2 x_1 +1) - t_2 (2 x_2+1) \right] \rho_n 
   \nonumber \\
&& \qquad \quad + {1 \over 8}  \left[ u_1 (2 + y_1) + u_2 (2+y_2) 
   \right] \rho_\Lambda 
\eeq
(The effective mass of the proton follows from replacing $\rho_n$
by $\rho_p$ in this expression) and
\beq
{\hbar^2 \over 2 m_\Lambda^*} =  {\hbar^2 \over 2 m_\Lambda} 
   + {1 \over 8} \left[ \lambda_1   + 3 \lambda_2 \right] \rho_\Lambda  
   +  {1 \over 8}  \left[ u_1 (2 + y_1) + u_2 (2+y_2) 
   \right] \rho_N 
\eeq

For deriving the Landau parameters we will also need the polarized energy 
functional  ${\cal E}(\rho_{n\uparrow}, \rho_{n\downarrow},$
$\rho_{p\uparrow}, \rho_{p\downarrow}, \rho_{\Lambda\uparrow}, 
\rho_{\Lambda\downarrow})$.
We do not reproduce this somewhat lengthy expression here, but note that it 
coincides for $np$ matter with the functional given by Bender {\it et al.} 
\cite{BDEN02}.

The Landau parameters in the monopolar approximation $\ell=0$ can be calculated
in the following way. We first take the derivative of the single particle 
energies (or equivalently the second functional derivative of the energy
density with respect to occupation numbers $\rho_i(k)$):
\beq
f_{\tau_1\sigma_1\tau_2\sigma_2} &:=&  \left( {d U_{\tau_1\sigma_1}(k) \over 
   d \rho_{\tau_2\sigma_2}} \right)_{| k=k_{F\tau_1\sigma_1}} \quad ;
   \quad f_{\tau_1\sigma_1\tau_2\sigma_2} =f_{\tau_2\sigma_2\tau_1\sigma_1} 
   \ , \quad \tau \in \left\{ n,p,\Lambda \right\} ,\ 
    \sigma \in \left\{ \uparrow,\downarrow \right\} \nonumber \\
\eeq
In the Skyrme model the single particle energies $U_{\tau\sigma}(k)$ are
quadratic in the momentum $k$
\beq
U_{\tau\sigma}(k) &:=& U_{\tau\sigma} + {\hbar^2 \over m^*_{\tau\sigma}} 
 k^2 
\eeq
and related to the chemical potentials in polarized matter
\beq
\mu_{\tau\sigma} &:=& \left({d {\cal E} \over d \rho_{\tau\sigma}}
   \right)_{| \rho{(\tau'\sigma' \not= \tau\sigma)} =cst} 
  = U_{\tau\sigma} +{\hbar^2 \over m^*_{\tau\sigma}} 
   k_{F\tau\sigma}^2 =  U_{\tau\sigma}(k_{F\tau\sigma})
\eeq
with $\rho_{\tau\sigma} = k_{F\tau\sigma}^3/(6 \pi^2)$ at $T=0$.

\vskip 0.2cm\noindent
The Landau parameters in the spin $S=0$ channel are obtained from
\beq
f_{\tau_1\tau_2} &:=& {1 \over 4} \left[ f_{\tau_1\uparrow\tau_2\uparrow} 
   + f_{\tau_1\downarrow\tau_2\downarrow} + f_{\tau_1\uparrow\tau_2\downarrow}
   + f_{\tau_1\downarrow\tau_2\uparrow} \right]_{|{\rm unpolarized}} 
\eeq
and in the spin $S=1$ channel 
\beq
g_{\tau_1\tau_2} &:=& {1 \over 4} \left[ f_{\tau_1\uparrow\tau_2\uparrow} 
   + f_{\tau_1\downarrow\tau_2\downarrow} - f_{\tau_1\uparrow\tau_2\downarrow}
   - f_{\tau_1\downarrow\tau_2\uparrow} \right]_{|{\rm unpolarized}}
\eeq
The $f_{\tau_1\tau_2}$ are related to usual thermodynamical quantities,
for example the compressibility and the asymmetry energy
\beq
K &=& {3 k_F^2 \over m_N^*} (1+F_0)\quad , \qquad
a_{\rm asym} = {k_F^2 \over 6 m_N^*} (1+F_0') \nonumber \\
{\rm with } && k_F=k_{Fn}=k_{Fp}\, ,\quad f_{nn}=f_{pp}\, , \quad
f_0={f_{pp}+f_{np}\over 2}\, ,\ f_0'={f_{pp}-f_{np}\over 2}\, , \nonumber \\
&& F_0=N_0\ f_0\, , \ F_0'=N_0\ f_0'\, , N_0={2 m_N^* k_F \over \pi^2}
\eeq 
in symmetric nuclear matter.

\vskip 0.2cm \noindent
The $g_{\tau_1\tau_2}$ are related to the magnetic susceptibilities
and are used to form the ferromagnetic criterion (Eqs. \ref{eq:Dij-gij}
-- \ref{eq:ferro})

\vskip 0.2cm \noindent
Their explicit expressions are 
\beq
\noalign{ \hskip -0.7cm
\underline{\mbox{In the spin $S=0$ channel}}
\vskip 0.2cm}
f_{nn} &=& {1 \over 2} t_0 (1 -x_0) +{1 \over 12} t_3 \rho_N^\alpha (1 -x_3) 
     +{1 \over 3} \alpha t_3 \rho_N^{\alpha-1} \left[ (1+{x_3 \over 2})
     \rho_N -({1 \over 2} + x_3) \rho_n \right] \nonumber \\
     && +{1 \over 12} \alpha (\alpha-1) t_3 \rho_N^{\alpha-2} \left[ 
     (1+{x_3 \over 2}) \rho_N^2 -({1 \over 2} + x_3) (\rho_n^2 + \rho_p^2) 
     \right] +{1 \over 4} \left[ t_1(1-x_1)+3 t_2(1+x_2) \right] k_{Fn}^2
     \nonumber \\
     && +{3 \over 8} u_3 (1+{y_3 \over 2}) \beta (\beta+1) \rho_N^{\beta-1} 
     \rho_\Lambda +{1 \over 4} \lambda_3 \gamma (\gamma-1) \rho_N^{\gamma-2} 
     \rho_\Lambda^2 \\
f_{np} &=&  t_0 (1 + {x_0 \over 2} ) +{1 \over 6} t_3 \rho_N^\alpha 
     (1 +{x_3 \over 2}) +{1 \over 4} \alpha t_3 \rho_N^\alpha \nonumber \\
      && +{1 \over 12} \alpha (\alpha-1) t_3 \rho_N^{\alpha-2} \left[ 
     (1+{x_3 \over 2}) \rho_N^2 -({1 \over 2} + x_3) (\rho_n^2 + \rho_p^2) 
     \right] \nonumber \\
     && +{1 \over 4} \left[ t_1(1+{x_1\over 2})+ t_2(1+{x_2\over 2}) 
     \right] (k_{Fn}^2 + k_{Fp}^2)
     \nonumber \\
     && +{3 \over 8} u_3 (1+{y_3 \over 2}) \beta (\beta+1) \rho_N^{\beta-1} 
     \rho_\Lambda +{1 \over 4} \lambda_3 \gamma (\gamma-1) \rho_N^{\gamma-2} 
     \rho_\Lambda^2 \\
f_{pp} &=& {1 \over 2} t_0 (1 -x_0) +{1 \over 12} t_3 \rho_N^\alpha (1 -x_3) 
     +{1 \over 3} \alpha t_3 \rho_N^{\alpha-1} \left[ (1+{x_3 \over 2})
     \rho_N -({1 \over 2} + x_3) \rho_p \right] \nonumber \\
     && +{1 \over 12} \alpha (\alpha-1) t_3 \rho_N^{\alpha-2} \left[ 
     (1+{x_3 \over 2}) \rho_N^2 -({1 \over 2} + x_3) (\rho_n^2 + \rho_p^2) 
     \right] +{1 \over 4} \left[ t_1(1-x_1)+3 t_2(1+x_2) \right] k_{Fp}^2
     \nonumber \\
     && +{3 \over 8} u_3 (1+{y_3 \over 2}) \beta (\beta+1) \rho_N^{\beta-1} 
     \rho_\Lambda +{1 \over 4} \lambda_3 \gamma (\gamma-1) \rho_N^{\gamma-2} 
     \rho_\Lambda^2 \\
f_{n\Lambda} &=& {1 \over 2} u_0 (2 + y_0) +{3 \over 16} u_3 (2+y_3) 
     (1+\beta) \rho_N^\beta + {1\over 8}  \left[u_1 (2+y_1) + u_2 (2+y_2) 
     \right] (k_{F\Lambda}^2 + k_{Fn}^2) \nonumber \\
     && +{1 \over 2} \lambda_3 \gamma \rho_N^{\gamma-1} \rho_\Lambda \\
f_{p\Lambda} &=& {1 \over 2} u_0 (2 + y_0) +{3 \over 16} u_3 (2+y_3) 
     (1+\beta) \rho_N^\beta + {1\over 8}  \left[u_1 (2+y_1) + u_2 (2+y_2) 
     \right] (k_{F\Lambda}^2 + k_{Fp}^2) \nonumber \\ 
     && +{1 \over 2} \lambda_3 \gamma  \rho_N^{\gamma-1} \rho_\Lambda \\
f_{\Lambda\Lambda} &=& {1 \over 2} \lambda_0 + {1 \over 2} \lambda_3 
     \rho_N^\gamma + {1 \over 4} \left[ \lambda_1 + 3 \lambda_2 \right] 
     k_{F\Lambda}^2 \\
\noalign{ \vskip 0.5cm \hskip -0.7cm
\underline{\mbox{In the spin $S=1$ channel}}
\vskip 0.2cm}
g_{nn} &=& {1 \over 2} t_0 (x_0 -1) +{1 \over 12} t_3 \rho_N^\alpha (x_3-1)
     +{1 \over 4} \left[ t_1(x_1-1)+ t_2(1+x_2) \right] k_{Fn}^2  \\
g_{np} &=& {1 \over 2} t_0 x_0 +{1 \over 12} t_3 x_3 \rho_N^\alpha 
      +{1 \over 8} \left[ t_1 x_1 +t_2 x_2 \right] (k_{Fn}^2 +k_{Fp}^2)\\
g_{pp} &=& {1 \over 2} t_0 (x_0 -1) +{1 \over 12} t_3 \rho_N^\alpha(x_3-1)
     +{1 \over 4} \left[ t_1(x_1-1)+ t_2(1+x_2) \right] k_{Fp}^2\\
g_{n\Lambda} &=& {1 \over 2} u_0 y_0 +{3 \over 16} u_3 y_3 \rho_N^\beta + 
{1 \over 8} \left[ u_1 y_1 + u_2 y_2 \right] (k_{F\Lambda}^2 + k_{Fn}^2) \\
g_{p\Lambda} &=& {1 \over 2} u_0 y_0 +{3 \over 16} u_3 y_3 \rho_N^\beta + 
{1 \over 8} \left[ u_1 y_1 + u_2 y_2 \right] (k_{F\Lambda}^2 + k_{Fp}^2) \\
g_{\Lambda\Lambda} &=& -{1 \over 2} \lambda_0 - {1 \over 2} \lambda_3 
     \rho_N^\gamma + {1 \over 4} \left[ - \lambda_1 +  \lambda_2 \right] 
     k_{F\Lambda}^2
\eeq

In the limit where hyperons are absent these expressions coincide with
those of Hern\'andez {\it et al.} \cite{HNP97}

\end{document}